\documentclass[a4paper,fleqn,usenatbib]{mnras}
\usepackage[T1]{fontenc}
\usepackage{ae,aecompl}

\usepackage{graphicx}	
\usepackage{amsmath}	
\usepackage{amssymb}	
\usepackage{subfig}
\usepackage{hyperref}

\usepackage{adjustbox}
\usepackage{color}
\usepackage{upgreek}
\usepackage{float}

\usepackage{booktabs}

\title[{\em{AstroSat}} observation of Mkn\,421]{Understanding the X-ray spectral curvature of Mkn\,421 using broadband {\em{AstroSat}} observations}

\author[Hota et al.]{Jyotishree Hota,$^1$\thanks{E-mail: hotajyoti@gmail.com}
Zahir Shah,$^{2,3}$\thanks{shahzahir4@gmail.com} 
Rukaiya Khatoon,$^4$
Ranjeev Misra,$^2$
\newauthor Ananta C. Pradhan,$^1$ 
and Rupjyoti Gogoi,$^4$
\\
$^1$ Department of Physics and Astronomy, National Institute of Technology Rourkela, Odisha ,India\\
$^2$ Inter-University Center for Astronomy and Astrophysics, Post Bag 4, Ganeshkhind, Pune - 411007, India\\
$^3$ Department of Physics, Central University of Kashmir, India \\
$^4$ Tezpur University,Napaam-784028, Assam, India.
}

\date{Accepted XXX. Received YYY; in original form ZZZ}

\pubyear{2021}
\begin{document}
\label{firstpage}
\pagerange{\pageref{firstpage}--\pageref{lastpage}}
\maketitle

\begin{abstract}
  We present a  time-resolved X-ray spectral study of the high energy peaked blazar Mkn\,421 using  simultaneous broadband observations from the LAXPC and SXT instruments on-board {\em{AstroSat}}. The $\sim 400$ ksec long observation taken during 3--8 January, 2017 was divided into segments of 10 ksecs. Each segment was fitted using synchrotron emission from  particles whose energy distribution was represented by a log-parabola model. We also considered particle energy distribution models where (i) the radiative cooling leads to a maximum energy ($\xi_{max}$ model), (ii) the system has energy dependent diffusion (EDD) and (iii) has energy dependent acceleration (EDA). We found that all these models describe the spectra, although the EDD and EDA models were marginally better. Time resolved spectral analysis allowed for studying the correlation between the spectral parameters for different models. In the simplest and direct approach, the observed correlations are not compatible with the predictions of the  $\xi_{max}$ model. While the EDD and EDA models do predict the correlations, the values of the inferred physical parameters are not compatible with the model assumptions. Thus, we show that spectrally degenerate models, can be distinguished based on spectral parameter correlations (especially those between the model normalization and spectral shape ones) making time-resolved spectroscopy a powerful tool to probe the nature of these systems.
\end{abstract}


\section{Introduction}

Blazars belong to a subclass of radio loud active galactic nuclei (AGN) with a relativistic jet directed towards the observer \citep{Urry_1995,2019Blandford}. They are characterized by non-thermal broadband spectra, rapid variability, strong radio and optical polarization 
\citep{2000Sambruna,Fan}. The  spectral  energy  distribution (SED) of a blazar extends from radio to $\gamma$-ray energies \citep{1998Fossati} and is  generally characterized  by a double hump structure with the low energy hump peaking in the optical to X-ray energies and the high energy hump with peak located in $\gamma$-ray band.
The high-energy emission can be produced either through leptonic models or through hadronic models. Based on leptonic model, there are two types of emission processes, that have been used to explain the broadband SED viz. synchrotron process and the inverse-Compton (IC) process. 
The low energy hump is produced by relativistic electrons interacting with ambient jet magnetic field through synchrotron process,  where as the high energy hump is mainly explained by IC scattering of low-energy photons \citep{1982Urry,1985Ghisellini,1987Begelman,1995Blandford,1996Bloom,2004Sokolov}.
The seed photons for IC scattering in  leptonic model can be either synchrotron photons (synchrotron self-Compton, (SSC): \citet{1974Jones,1992Maraschi,1993Ghisellini}) or photons external to jet (external Compton (EC): \citet{1992Dermer,1994Sikora,2000Bazejowski,2017Shah}).
Alternatively, in the hadronic processes, the the higher frequency emission can also be initiated by relativistic protons such as the proton synchrotron process and pion production. \citep{1992Mannheim,2001Mucke}.

Blazars are broadly classified into Bl Lacs and FSRQs. Additionally, based on the location of the peak of synchrotron component, BL Lacs are further classified into low-energy peaked BL Lacs (LBL), ($\nu_p$ < $10^{14}$ Hz), Intermediate peaked BL lacs (IBL) ($10^{14}$ < $\nu_p$ < $10^{15}$) and high-energy peaked BL Lacs (HBL) ($\nu_p$ > $10^{15}$) \citep{1995ApJ...444..567P}. 

The Blazar Mkn\,421 is a well-known nearby HBL source, which is located at a redshift of z = 0.031 \citep{Punch1992}. It shows rapid variability in a time scale of few minutes to days. Mkn\,421 has been observed across the electromagnetic spectrum by various advanced observatories  in order to explore its broadband SED characteristics and underlying physical processes associated  
\citep{2001Krawczynski,2005Brinkmann,2007Albert,2008Lichti,2008Fossati,Acciari_2009,2009Tramacere,2009Horan,Ushio_2009,Ushio2010fi,2010Aleksi,2010Isobe,2012Aleksi,Shukla,2016Bartoli,2016Sinha}.

 Mkn\,421 is an extremely bright source and displays a strong variability in all wavelengths from radio to very high energy gamma rays on multiple time scales \citep{Wagner,2014Falomo,2020Goyal}. Its long-term multi-wavelength study gives clues on the emission mechanisms, structure and physical parameters inside the relativistic jets. 
Several previous studies have reported correlated variability between the X-ray and TeV $\gamma$-ray emission from Mkn\,421 during flaring period
\citep{1995Macomb,1996Buckley,Fossati2004ar,2007Albert,2008Fossati,2009Donnarumma,Abdo_2011,2011Acciari,cao,2015Aleksicb} 
as well as during low activity period \citep{2015Aleksica,Balokovic}. 
Where as no significant correlation between optical/UV and X-ray/Tev emission has been reported \citep{1995Macomb,2007Albert,cao}.

In its flaring state the X-ray spectra of Mkn\,421 shows significant curvature  \citep{2000Fossati,2004Massaro}, and the log-parabola model has been shown to provide the satisfactorily fit to such spectral shapes \citep{2004Massaro,2004A&A...413..489M,2006A&A...448..861M,2009Donnarumma,2015Sinha}.
The broad band X-ray spectrum of Mkn\,421 was also analyzed by \citet{Ushio2010fi} with the parametric forward-fitting SYNCHROTRON model incorporated in XSPEC. The authors showed that the synchrotron emission from exponentially moderated power-law electron distribution provided a good fit to the X-ray spectrum \citep{Ushio2010fi}.
Detailed correlation study among the best fit model parameters has been carried out in different studies \citep{2007Tramacere, 2008A&A...478..395M, 2009Tramacere,2011Tramacere, 2015Sinha}.
 Further, \citet{2020Goswami} showed that an analytical model obtained by considering the energy dependent electron escape time scale,  provides a good fit to the hard curved hard X-ray spectra of Mkn\,421.

Recently, \citet{2018ApJ...859L..21C} have analyzed the X-ray light curves of Mkn\,421 by using the observations made with SXT and LAXPC telescopes on board the {\em {\em{AstroSat}}} satellite. They showed that the X-ray power spectral density (PSD) is a piece-wise power-law with a break. The strong variability observed during this long observation also makes it an ideal one to study spectral variability as a function of time. In this work, we undertake a time resolved spectral study   using different particle energy distribution models. The motivation is to find out how well these models represent the observed spectra, obtain  correlations between spectral parameters and use these correlations to check for self consistency of the model.

 The paper is organised as follows: in \autoref{sec:astrosat}, we give details of the {\em{AstroSat}} observation and describe the data analysis procedure. In \autoref{sec:spectral models}, we present details of the spectral analysis of the observed data. In section \autoref{sec:discus} the results are summarized and their implications are discussed.


\section{{\em{AstroSat}} X-RAY OBSERVATIONS AND ANALYSIS}
\label{sec:astrosat}
 {\em {\em {\em {\em {\em AstroSat}}}}} is India's first multi-wavelength space astronomy satellite. It has five scientific instruments onboard  which carry observation of  a cosmic source over a wide range of energies from UV to hard X-ray \citep{AGRAWAL20062989,KulinderPalSingh,2016arXiv160806051R}. 
 The  onboard  instruments like Ultra-Violet Imaging Telescope (UVIT (130--300 nm ); \cite{2017JApA...38...28T, 2017AJ....154..128T}), Soft X-ray focusing Telescope (SXT (0.3--8.0 keV); \cite{2017JApA...38...29S, 2016SPIE.9905E..1ES}),


 Large Area X-ray Proportional Counter (LAXPC (3 -- 80 keV); \cite{2016SPIE.9905E..1DY}) and Cadmium Zinc Telluride Imager (CZTI (10 -- 100 keV); \cite{2017arXiv171010773R}), 
 allow for simultaneous observations in UV, soft X-rays and hard X-rays. {\em{AstroSat}} has the unique capability of probing the strictly simultaneous spectral curvature over broad energy ranges (X-ray and UV). During 3 -- 8 January, 2017, {\em{AstroSat}} carried a ToO observation of the flaring activity of Mkn\,421 for a total observation time of 409 ksec \citep{2018ApJ...859L..21C}. The data of this observation is available on the {\em{AstroSat}} data archive astrobrowse \footnote{https://astrobrowse.issdc.gov.in/astro\-archive/archive/Home.jsp}. The information of instrument-specific software/tools for processing the data is provided in the ASSC website \footnote{http://astrosat-ssc.iucaa.in/?q=data and analysis}. The details of the data reduction of SXT and LAXPC observations are as follows.

\subsection{SXT}
\label{sec:SXT} 
SXT is a X-ray focusing telescope which operates in the 0.3 -- 8.0 keV energy band \citep{bf149366599c4681bceaadea3bdf114d}. It is used for X-ray imaging and spectroscopy purposes with angular resolution of 2$'$ and FOV of $\sim$ 40$'$ diameter. The SXT observed Mkn\,421 in the Photon Counting (PC) mode in 57 orbits during the period 3 -- 8 January, 2017.
We processed the archived Level-1 data of each orbit with the SXT pipeline (AS1SXTLevel2,version 1.4b \footnote{http://www.tifr.res.in/$^\sim$astrosat$\_$sxt/sxtpipeline.html}) and obtained cleaned Level-2 event file for each orbit. In order to avoid the time-overlapping of events from the consecutive orbits, we merged the 
cleaned Level-2 event files from all the orbits into a single cleaned event file by using the standard julia script, developed by the instrument team.
The science products were generated from the merged event files using XSELECT (V2.4d) package which is in-built in HEASOFT. A circular region of 15$'$ radius centered at the source location was used to extract the source count rate, the 15$'$ source region was selected so that it includes more than 90$\%$ of the source photons.  The spectral analysis was carried out by using a background spectrum ``SkyBkg$\_$comb$\_$EL3p5$\_$Cl$\_$Rd16p0$\_$v01.pha" provided by the SXT POC team and the  response  matrix file  (RMF) (``sxt$\_$pc$\_$mat$\_$g0to12.rmf") and  an  off-axis  auxiliary  response  file  (ARF)  generated using SXT ARF generation tool\footnote{http://www.tifr.res.in/$^\sim$astrosat$\_$sxt/dataanalysis.html}. The grppha tool was used to ensure a minimum of 40 counts per bin. Due to uncertainities in the response at low energies, we restricted the spectra analysis to 0.5--7.0 keV.
Further, we add a systematic uncertainty of 3\% to the data in order to account for uncertainties in response calibrations. 
 Gain variation was taken into account using the  gain fit tool in Xspec with the slope fixed at one. The best fit offset value for different segments  was found to be typically $\sim 0.03$ keV.

\subsection{LAXPC}
\label{sec:LAXPC} 
The LAXPC is one of the primary instruments of {\it {\em{AstroSat}} } which consists of three non-imaging proportional counters namely LAXPC10, LAXPC20, LAXPC30. Each proportional counter are having an effective area
of $\sim$ 2000 $cm^2$. The LAXPC works in the energy range 3 -- 80 keV with high time resolution ($\sim$ 10 $\mu$s) \citep{J.S.Yadav,2017Antia,2017Agrawal,Misra_2017}. 
We processed the Level-1 data of the source with the LAXPCSOFT package which is based on Fortran codes and  available at the {\em{AstroSat}} Science Support Cell (ASSC) \footnote{http://astrosat-ssc.iucaa.in}. The command laxpc\_make\_event were used to create the combined level-2 event file and laxpc\_make\_stdgti were used to generate good time interval (gti) file in which the intervals of South Atlantic Anomaly (SAA) and Earth occultation were removed.
Finally the tasks laxpc$\_$make$\_$lightcurve and laxpc$\_$make$\_$spectra were used to generate the source lightcurve and spectra respectively, from the gti file.
The corrected light curve has not been background subtracted.  For background estimation we have used the faint source method \citep{2021arXiv210206402M}. The background subtraction has done separately by using the standard FTool lcmath command.
We limited the spectral analysis to 4 - 18 keV as we found that the background completely dominated the spectra at higher energies. Systematic uncertainty of 3\% was included in the spectral analysis. Only LAXPC 10 and LAXPC 20 were used for the analysis, since LAXPC 30 had substantial gas leak by the time of the observation.


\subsection{TIME VARIABILITY}
\label{sec:timevar}
\begin{figure}
	\includegraphics[width=0.7\columnwidth,angle=270  ]{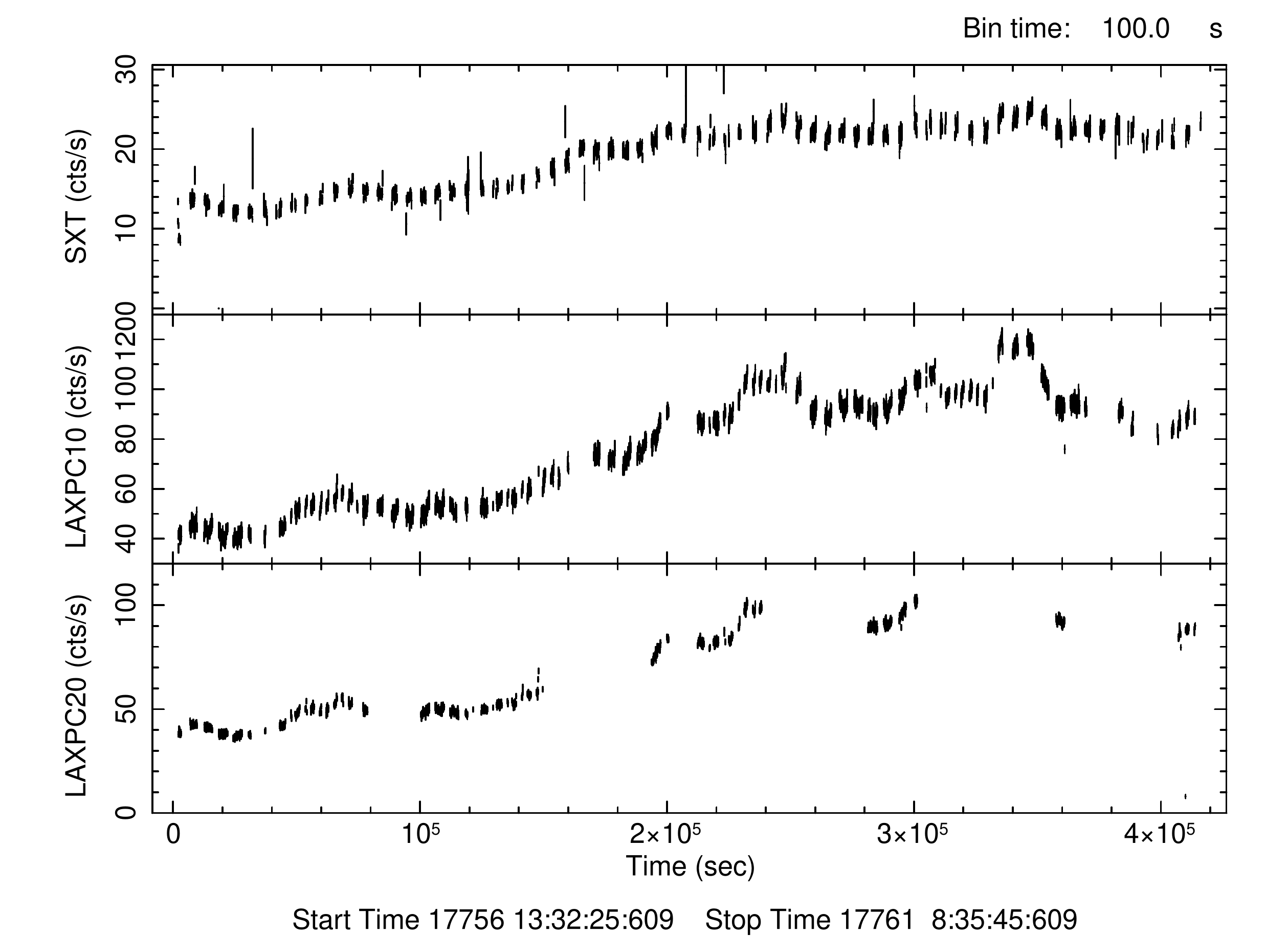}
   \caption{Light curve of Mkn\,421 in the energy range of 0.5--7.0 keV from SXT (top panel) and 3.0-30.0 keV from LAXPC10 (middle panel) and LAXPC20 (bottom panel).}
    \label{fig:light curve}
\end{figure}

The SXT light curve is generated for the time bin of 100 sec in the energy range 0.5 -- 7 keV (top panel of \autoref{fig:light curve}) where as the LAXPC10 and LAXPC 20 light curve are obtained for the time bin of 100 sec in the energy range 3 -- 30 keV (middle and bottom panels of \autoref{fig:light curve}).
 The AstroSat observations of Mkn\,421 during 3 -- 8 January, 2017 revealed that the source was in high flux state and undergoing through strong variablity in soft and hard X-rays (see \autoref{fig:light curve}). The calculated variability amplitude ($F_{var}$) of the source for SXT, LAXPC10, LAXPC20 using LCSTATS are 0.21772 $ \pm$ 0.004, 0.30334  $\pm$ 0.005, 0.37697 $\pm$ 0.009, respectively.
 

\section{ Time-resolved spectral analysis}
\label{sec:spectral models}

The X-ray emission in Mkn\,421 is known to be synchrotron emission produced by the interaction of non-thermal relativistic electrons with an ambient magnetic field. We  assume that the emission arises from a spherical region of radius, R which is filled with tangled magnetic field, B and relativistic isotropic electron distribution, $n(\gamma)$. If we express the electron lorentz factor, $\gamma$ in terms of $\xi$ such that $\xi=\gamma\sqrt{\mathbb{C}}$, where $\mathbb{C}=1.36 \times 10^{-11}\frac{\delta B}{1+z}$ with z being the redshift of source and $\delta$ as jet Doppler factor,  then the synchrotron flux received by the observer at energy $\epsilon$ will be  \citep{Begelman},\\
\begin{align}\label{eq:syn_conv}
	F_{\rm syn}(\epsilon) 
	= \frac{\delta^3(1+z)}{d_L^2} V \,\mathbb{A} \int_{\xi_{min}}^{\xi_{max}} f(\epsilon/\xi^2)n(\xi)d\xi
\end{align}
where, $d_L$ is the luminosity distance, V is the volume in the emission region,  $\mathbb{A}=\frac{\sqrt{3}\pi e^3B}{16m_ec^2\sqrt{\mathbb{C}}}$ and f(x) is the synchrotron emmisivity function \citep{1986Rybicki}. We solved  \autoref{eq:syn_conv} numerically and included it as a local convolution model, \emph{$synconv \otimes n(\xi)$} in  XSPEC. This allows for the photon spectrum to be modelled for any particle energy distribution $n (\xi)$ represented by any XSPEC model. 
It is important to mention here that XSPEC ``energy" variable in the \emph{$synconv \otimes n(\xi)$} model should be interpreted as $\xi=\gamma\sqrt{\mathbb{C}}$. 
Thus, any parameter of the model, such as the pivot energy of the logparabola model $\xi_{piv}$, would be related to $\gamma$ as $\gamma_{piv} = \xi_{piv}/\sqrt{\mathbb{C}}$. In order to account for Galactic absorption, we used the TBabs model \citep{2000ApJ...542..914W} available in the XSPEC with the  equivalent hydrogen column density ($N_{H}$)  fixed at  $1.92 \times 10^{20}$cm$^{-2}$ as obtained from the online tool{\footnote{https://heasarc.gsfc.nasa.gov/cgi-bin/Tools/w3nh/w3nh.pl}} which is developed by LAB survey group \citep{2005A&A...440..775K}.

\begin{figure*}
	\includegraphics[width=0.95\columnwidth]{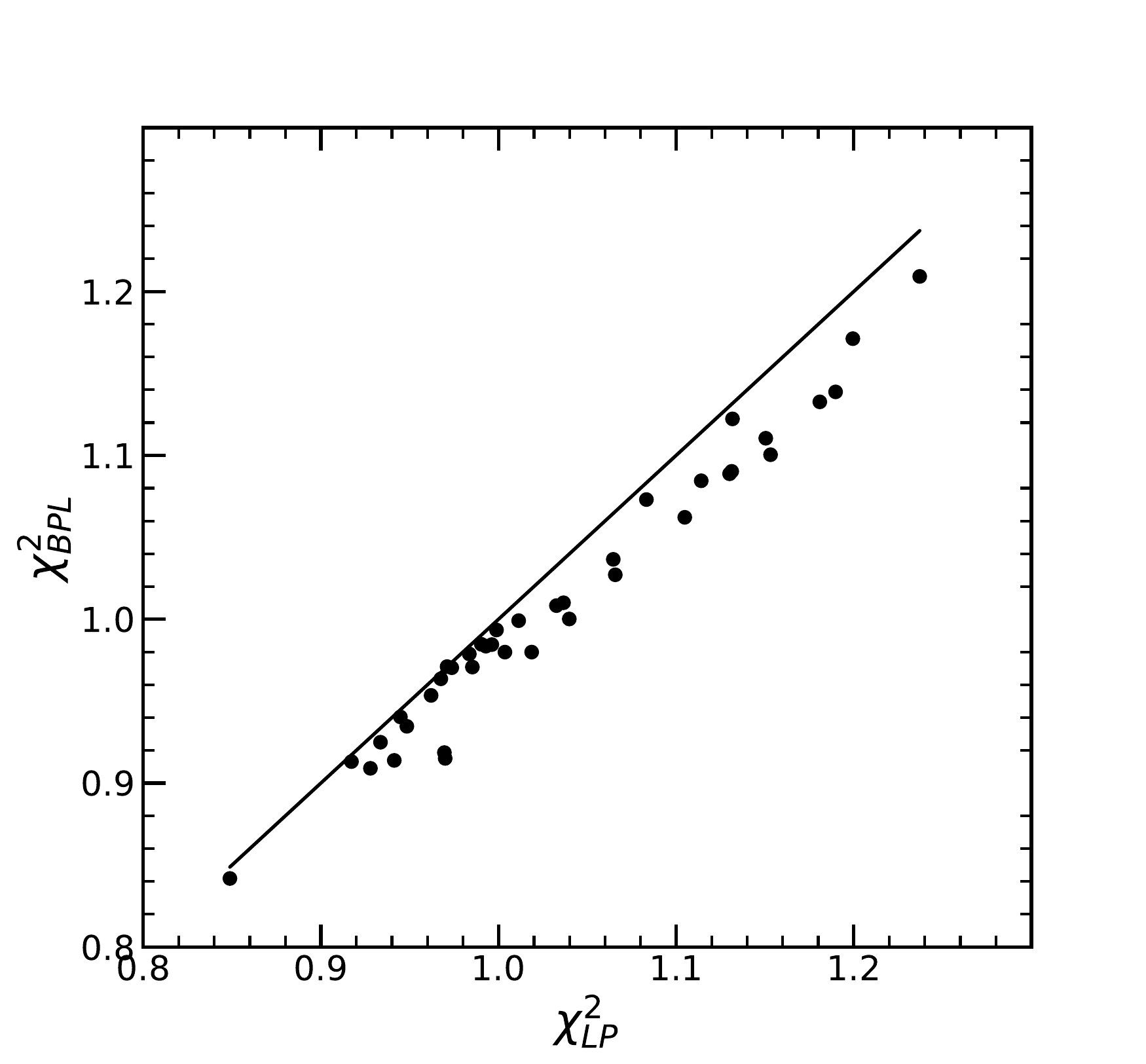}
	\includegraphics[width=0.95\columnwidth]{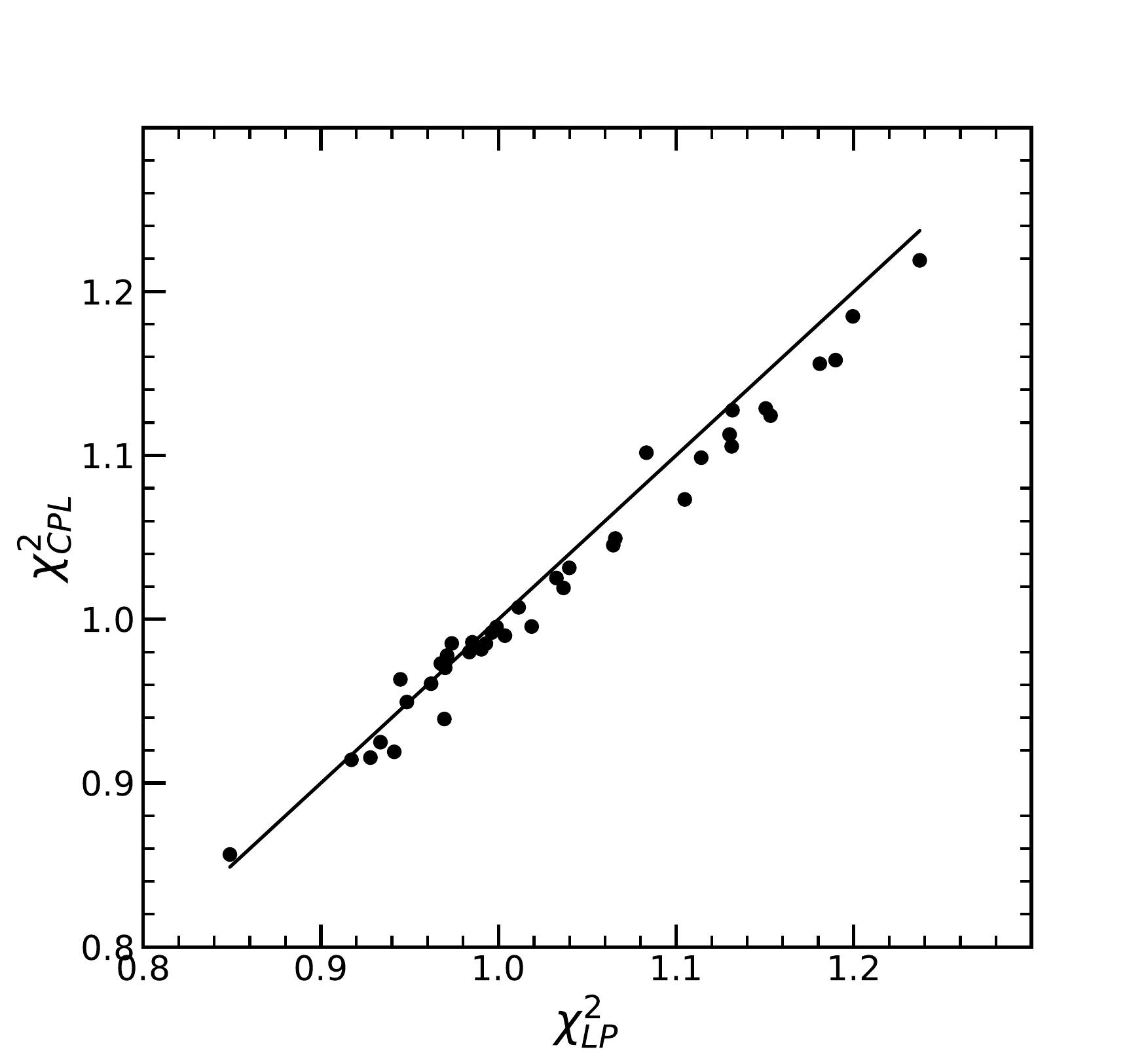}
\includegraphics[width=0.95\columnwidth]{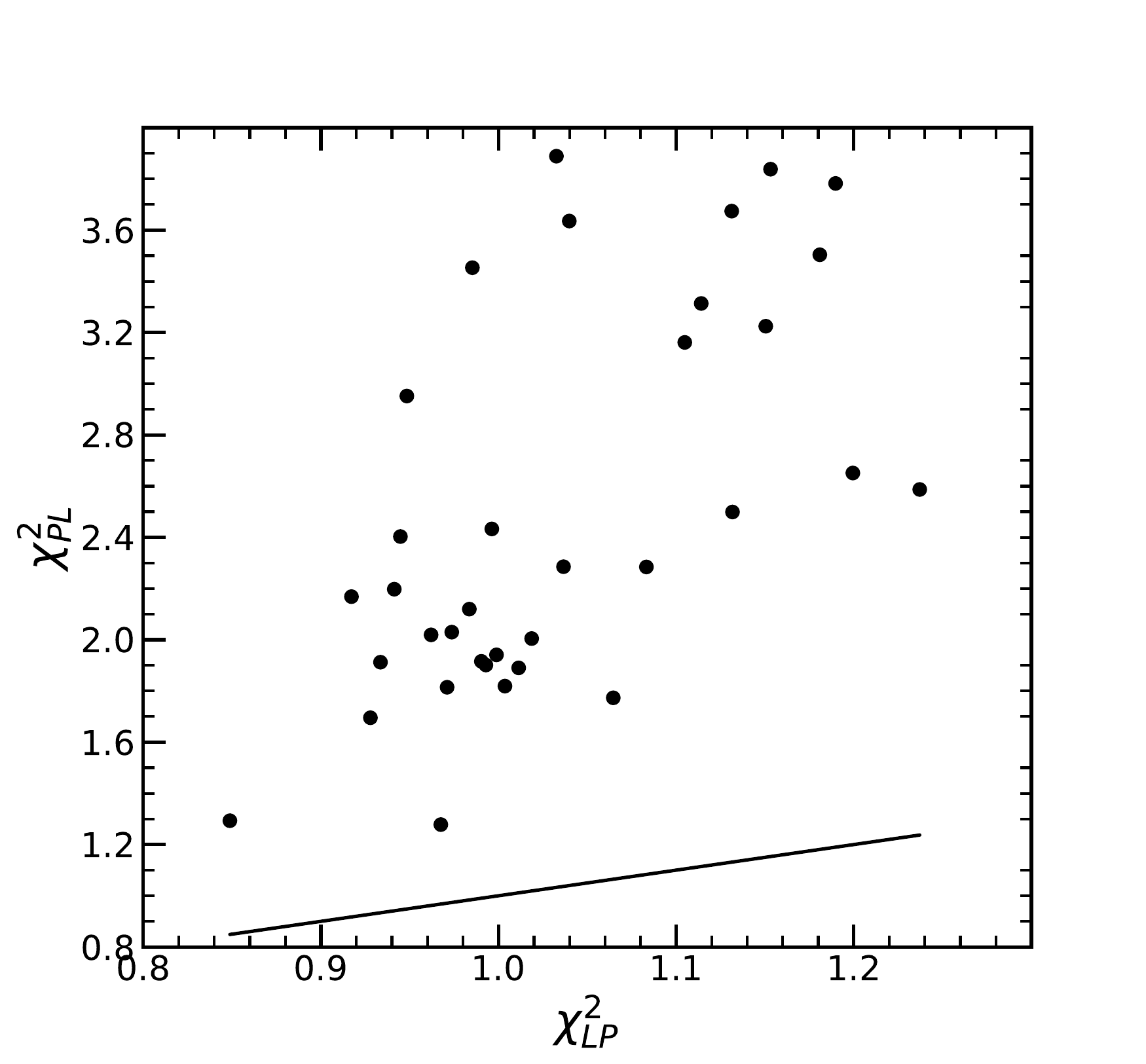}
    \caption{ Comparison of reduced-$\chi^{2}$ of synchrotron convolved logparabola model with the reduced $\chi^2$ of synchrotron convolved broken power law model (top left panel), the reduced $\chi^2$ of synchrotron convolved Power law with an exponential cutoff model (top right panel) and the reduced $\chi^2$ of synchrotron convolved PL model (bottom panel). In each panel, the solid line is identity line.
    }
    \label{fig:lpchisq}
\end{figure*}

The broadband X-ray spectrum of Mkn\,421 is  generally characterized by a smooth broad curvature (\citep{2000Fossati,2004Tanihata,2004Massaro,2007Tramacere,2007A&A...467..501T,2009Tramacere}), which can be obtained by considering the input particle density to be curved. There are several possible curved models which may give acceptable fits to the curved spectrum. Here, we fitted the observed X-ray spectrum in each 10 ksec time segment by synchrotron emission from a particle energy distribution described by empirical models viz power law (PL), logparabola (LP), broken power law (BPL), and power law with exponential cutoff (CPL). We have also fitted the observed spectrum with synchrotron emission from physical models such as power-law with a maximum energy due to radiative cooling, energy dependent diffusion (EDD), and energy dependent acceleration (EDA). Our results show that the empirical models (LP, BPL, CPL) fit the observed spectrum well (see top panel of \autoref{fig:lpchisq}), 
while power law model gives a poorer fit as compared to other models (bottom panel of \autoref{fig:lpchisq}). 
The motivation of our work is to obtain a more consistent physical scenario responsible for the emission, hence, we have discussed only physical models in detail.


\subsubsection{Logparabola model}

\begin{figure*}
 	\includegraphics[height=7.5cm, width=0.95\columnwidth]{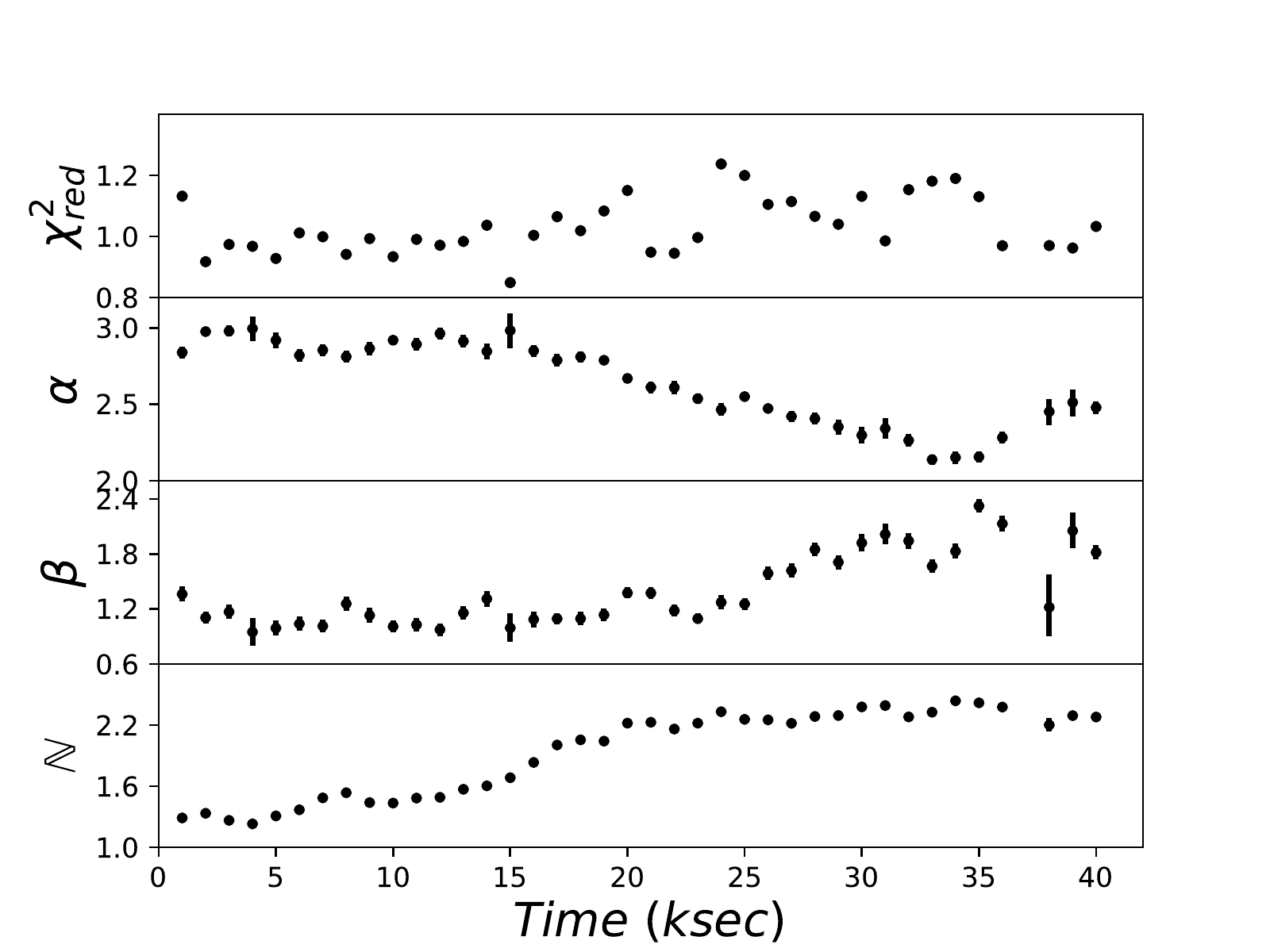}
	\includegraphics[width=0.95\columnwidth]{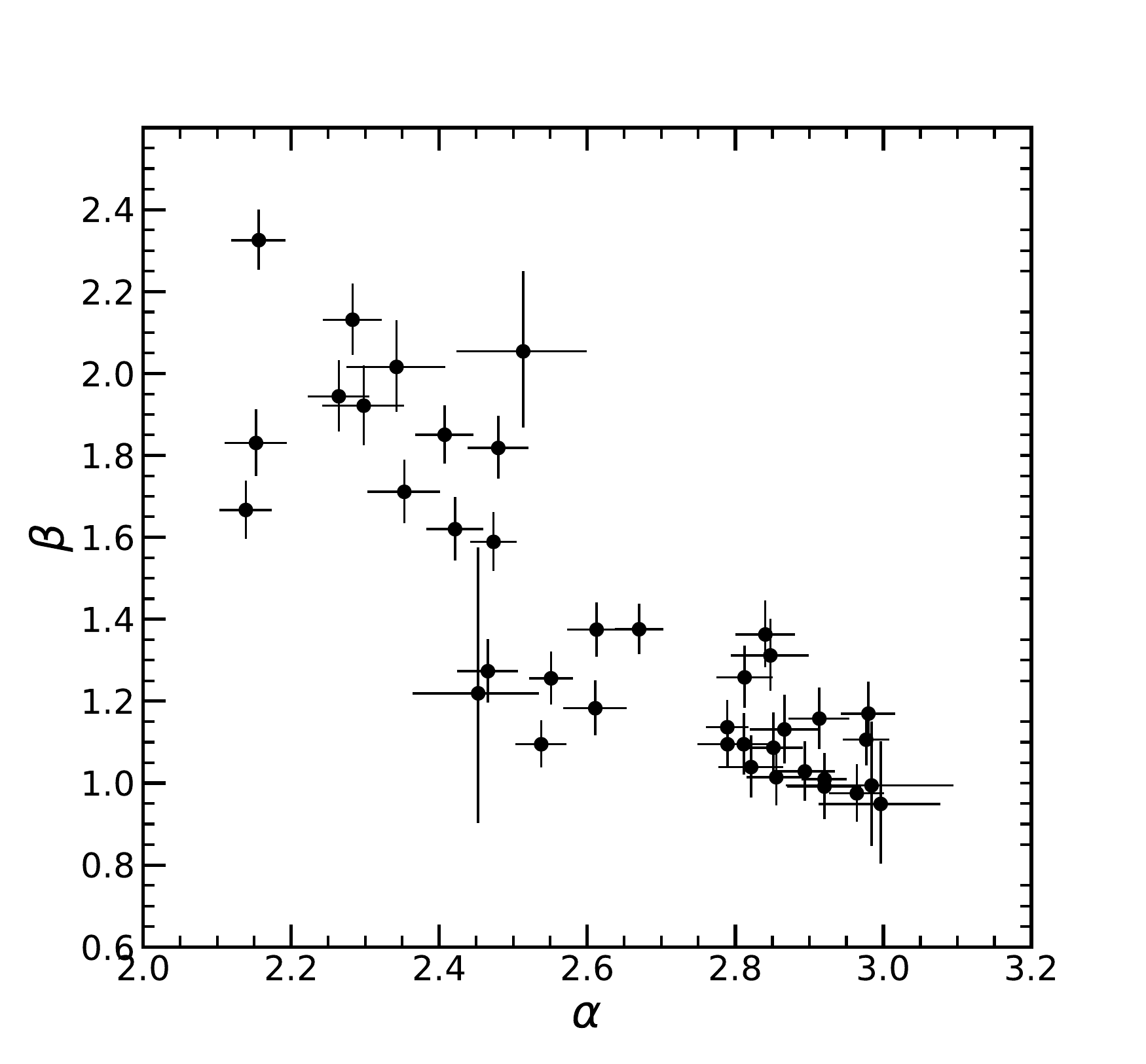}
	\includegraphics[width=0.95\columnwidth]{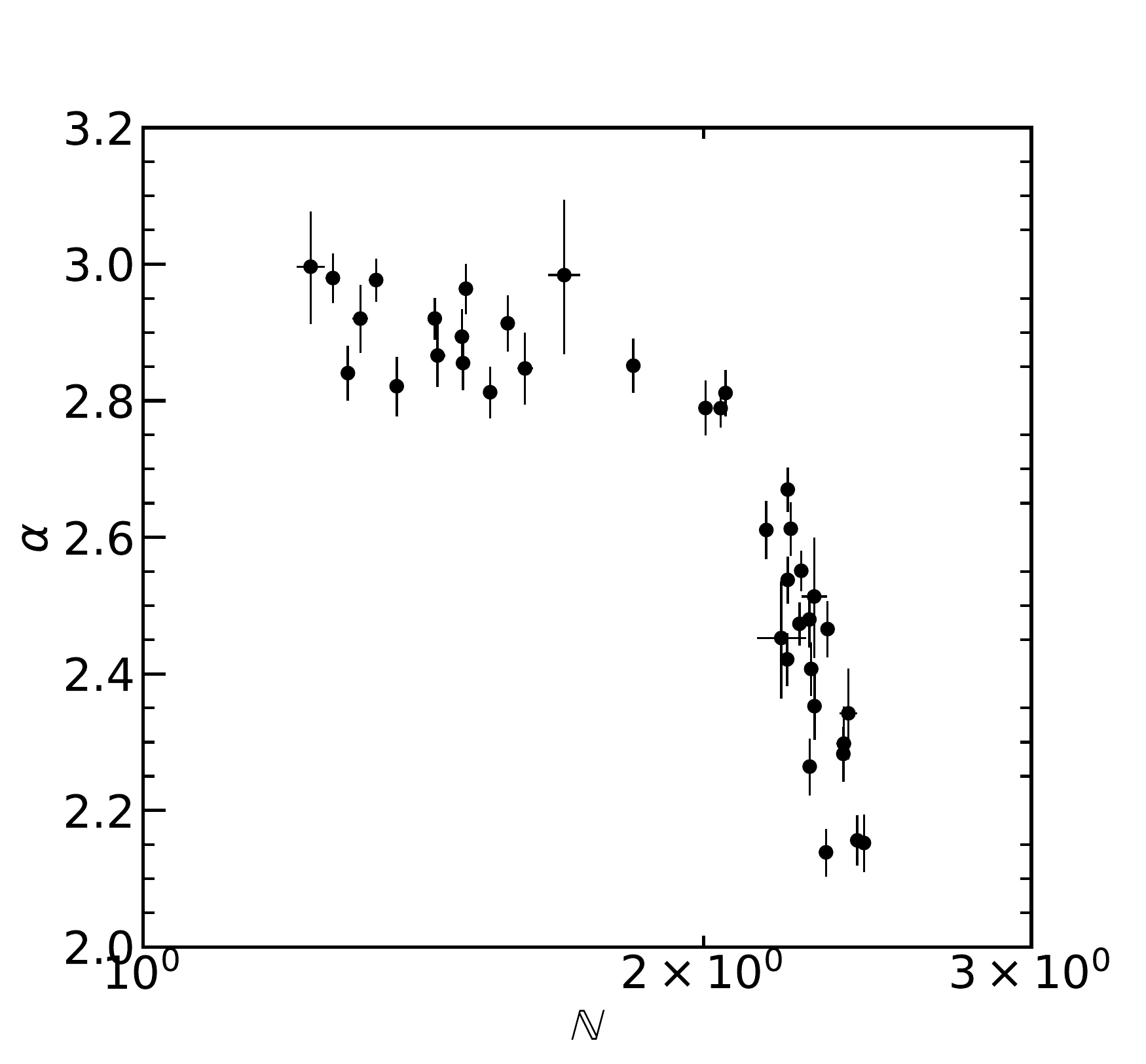}
	\includegraphics[width=0.95\columnwidth]{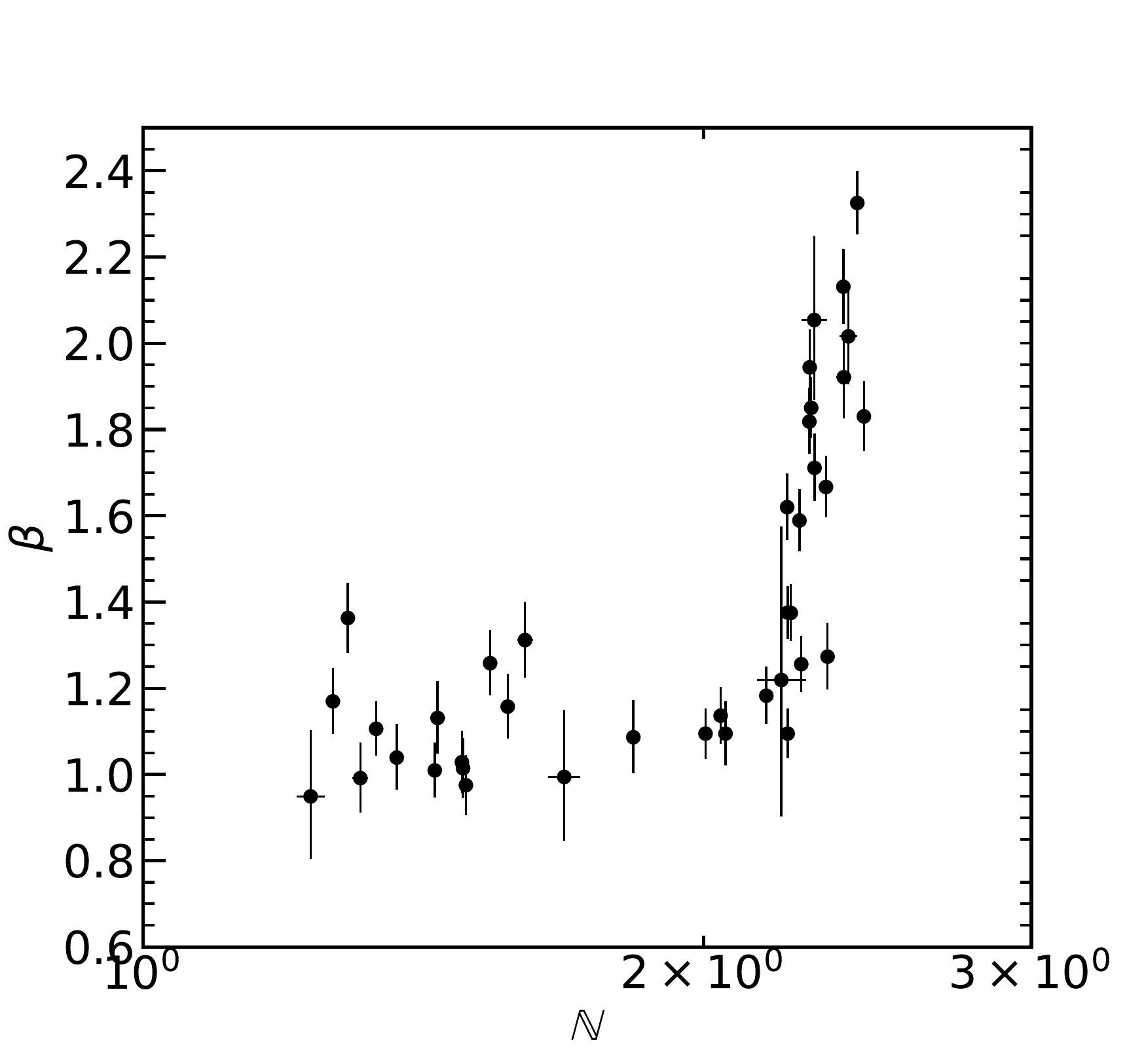}
	\includegraphics[width=0.95\columnwidth]{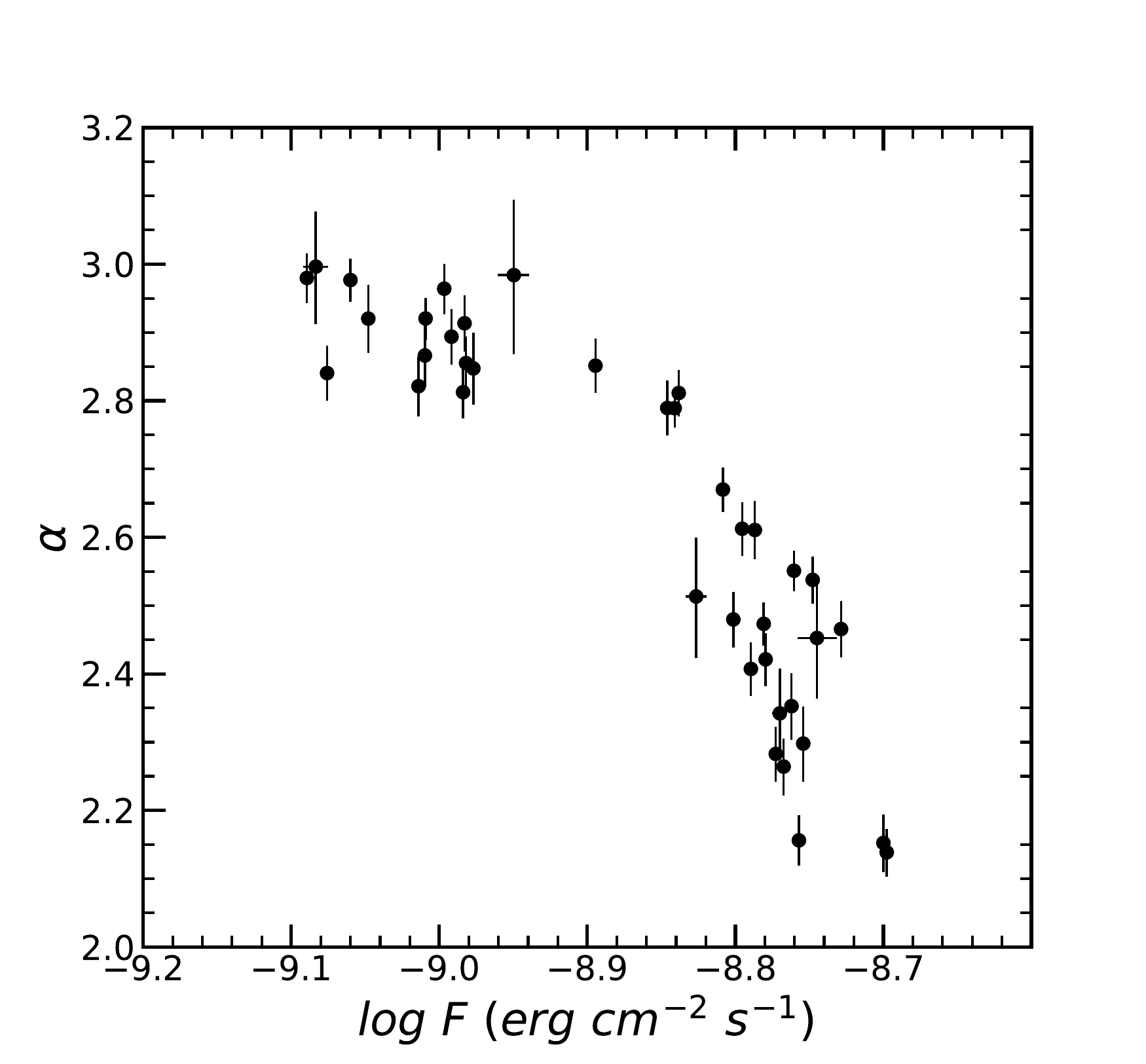}
	\includegraphics[width=0.95\columnwidth]{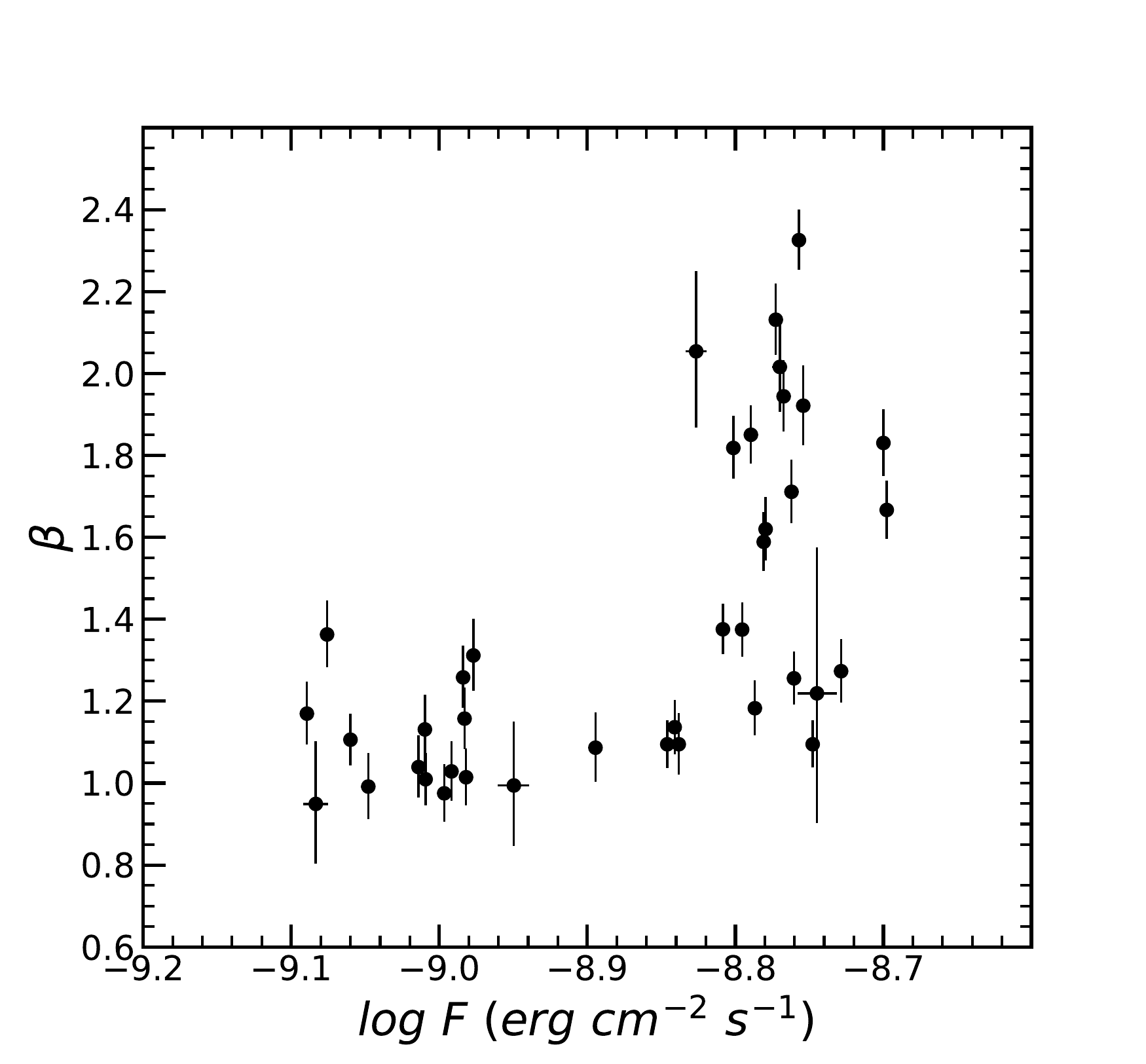}
    \caption{Top left panel: Best fit results obtained by fitting the wideband spectra in each time segment with synchrotron convolved logparabola model. Top right panel: scatter plot between best fitted $\alpha$ and $\beta$ values,  middle left panel: scatter plot between best fitted $\mathbb{N}$ and $\alpha$ values, middle right panel: scatter plot between best fitted $\mathbb{N}$ and $\beta$ values,
   bottom left panel: scatter plot between flux, $F_{0.5-18.0 \ kev}$ and best fitted $\alpha$ values, bottom right panel: scatter plot between flux, $F_{0.5-18.0 \ kev}$ and best fitted $\beta$ values.
    }
    \label{fig:alpha_betaflux}
\end{figure*}

We first fitted the broadband X-ray spectra using \emph{$synconv \otimes n(\xi)$}, where $n(\xi)$ is given by a logparabola form i.e.,
\begin{equation}
 \label{eq:logpar_conv}
    n(\xi)  = K ({{\xi}/{\xi_{r}}})^{-{\alpha}-{\beta}{\log({{\xi}/{\xi_{r}}})}}\\
\end{equation}
Here, ${\alpha}$ is the particle spectral index at the reference energy, $\xi^2=\xi_r^2$,  ${\beta}$ is the curvature parameter and K is the normalization of the particle density at $\xi=\xi_r$. While performing the 
 spectral fit in each time bin with \emph{$synconv \otimes n(\xi)$} model (Equation \ref{eq:syn_conv}), we have fixed the $\xi_r^2$  at 1 keV, and the three parameters, viz. ${\alpha}$, ${\beta}$ and norm $\mathbb{N}$ were kept free. Here  $\mathbb{N}$ is defined as
\begin{equation}
 \label{eq:norm_lp}
   \mathbb{N}  = \frac{\delta^3(1+z)}{d_L^2} V \,\mathbb{A}K
\end{equation}
The model provides reasonable reduced-$\chi^2$ for all the time segments, as shown in top left panel of \autoref{fig:alpha_betaflux} where the reduced $\chi^2$ and  the best-fit parameters are plotted as a function of time.
The rest of the panels show the correlation between the different parameters.
In \autoref{Tab:loggammamax} (left), we provide the best-fit parameters for all the time segment and the reduced $\chi^{2}$ value using logparabola model.
 We note that  $\beta$ and $\alpha$ are clearly inversely correlated. The value of $\alpha$ decreases slowly with normalization till $\mathbb{N} \sim 2$ and then decreases more rapidly, while for $\beta$ there is a slow increase followed by a rapid increase.
 A strong anti-correlation is obtained between $\alpha$ and flux, $F_{0.5-18.0 \ kev}$ with r-value and p-value as  $\sim -0.89$  and $1.6 \times 10^{-14}$, respectively, which is hardening when brightening features commonly seen in blazars (left bottom panel of \autoref{fig:alpha_betaflux}) . Whereas a positive correlation is observed between $\beta$ and flux, $F_{0.5-18.0 \ kev}$ with r-value and p-value $\sim 0.64$ and $1.6 \times10^{-5}$, respectively (right bottom panel of \autoref{fig:alpha_betaflux}). 

\subsubsection{Particle distribution with a maximum energy}
We next consider the case of particles getting accelerated by a shock and then subsequently these accelerated particles lose energy by emitting radiation through the synchrotron process. Since the synchrotron radiative loss is proportional to  energy squared, it is more significant for higher energies,
leading to a maximum energy for the distribution.  The steady state particle density for such a scenario is determined by
\begin{equation}\label{kinetic}
\frac{\partial}{\partial \gamma}\left[\left(\frac{\gamma} {\tau_{acc}}-\beta_s\gamma^2\right)n_a\right]+\frac{n_a}{\tau_{esc}}=Q\delta(\gamma-\gamma_0)
\end{equation}
where $\tau_{acc}$ and $\tau_{esc}$ are the acceleration and escape time-scales respectively, and the cooling term contains 
$\beta_s=\frac{4}{3}\frac{\sigma_TB^2}{8\pi m_ec}$, with B as the magnetic field, $\sigma_T$ the  Thomson cross section, and $m_e$ the electron mass.  It is assumed that there is mono-energetic injection of particles at energy $\gamma_0$. The solution to the above equation (after transforming $\gamma$ to $\xi = \sqrt{\mathbb{C}} \gamma$) is \citep{1962SvA.....6..317K},
\begin{equation}
  \label{eq:gmax_part}
  n(\xi) = K \xi^{-p} \left(1-\frac{\xi}{\xi_{max}}\right)^{(p-2)}
\end{equation}
where $K=Q_0\tau_a\gamma_0^{p-1}\mathbb{C}^{p/2}$,   $p= \tau_{acc}/\tau_{esc} +1$  is the particle spectral index,  $\xi_{max}=\gamma_{max}\sqrt{\mathbb{C}}$ and
$\gamma_{max}  = 1/(\beta_s \tau_{acc})$.

\begin{figure*}
	\includegraphics[width=0.85\columnwidth]{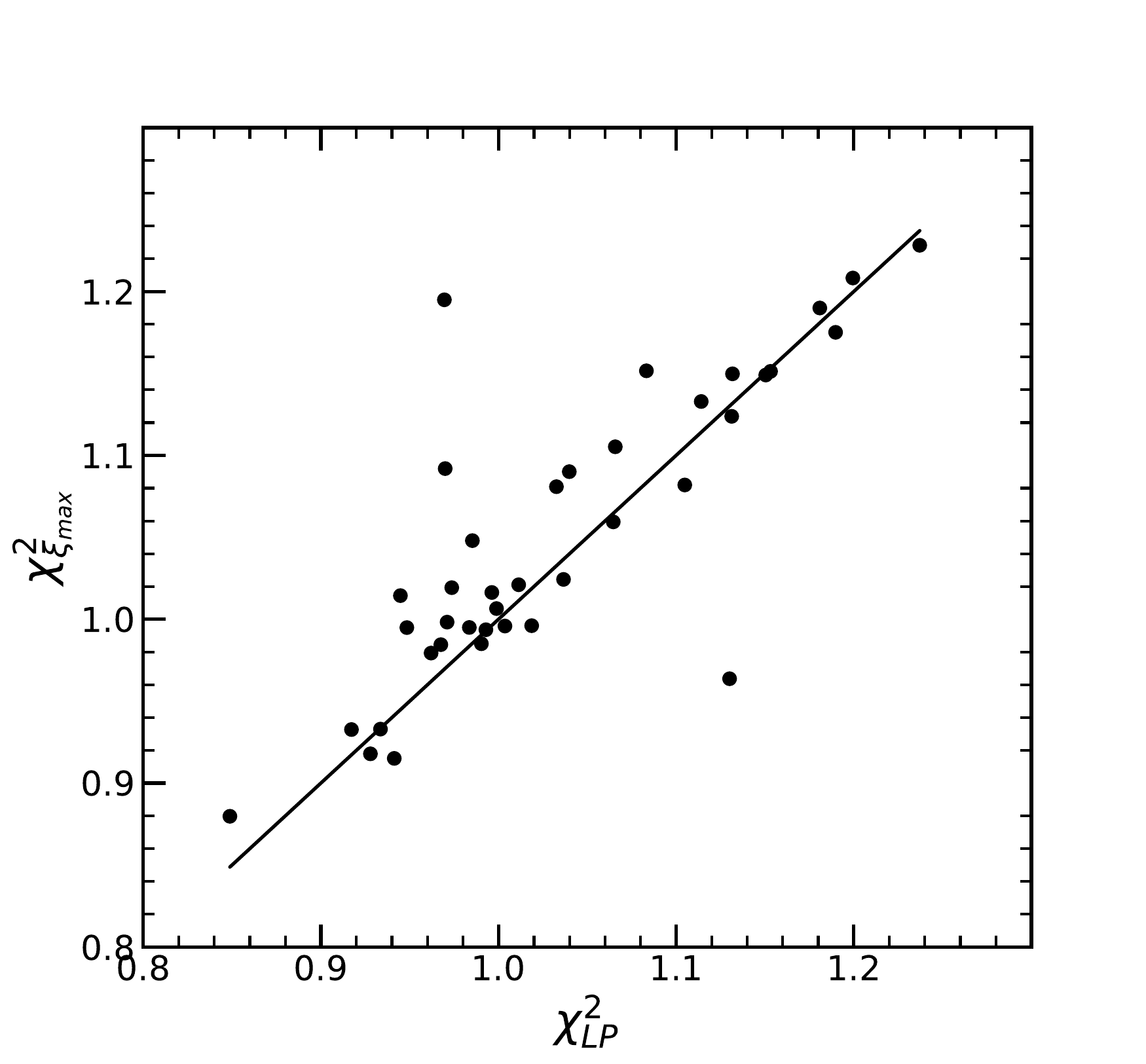}
	\includegraphics[width=0.85\columnwidth]{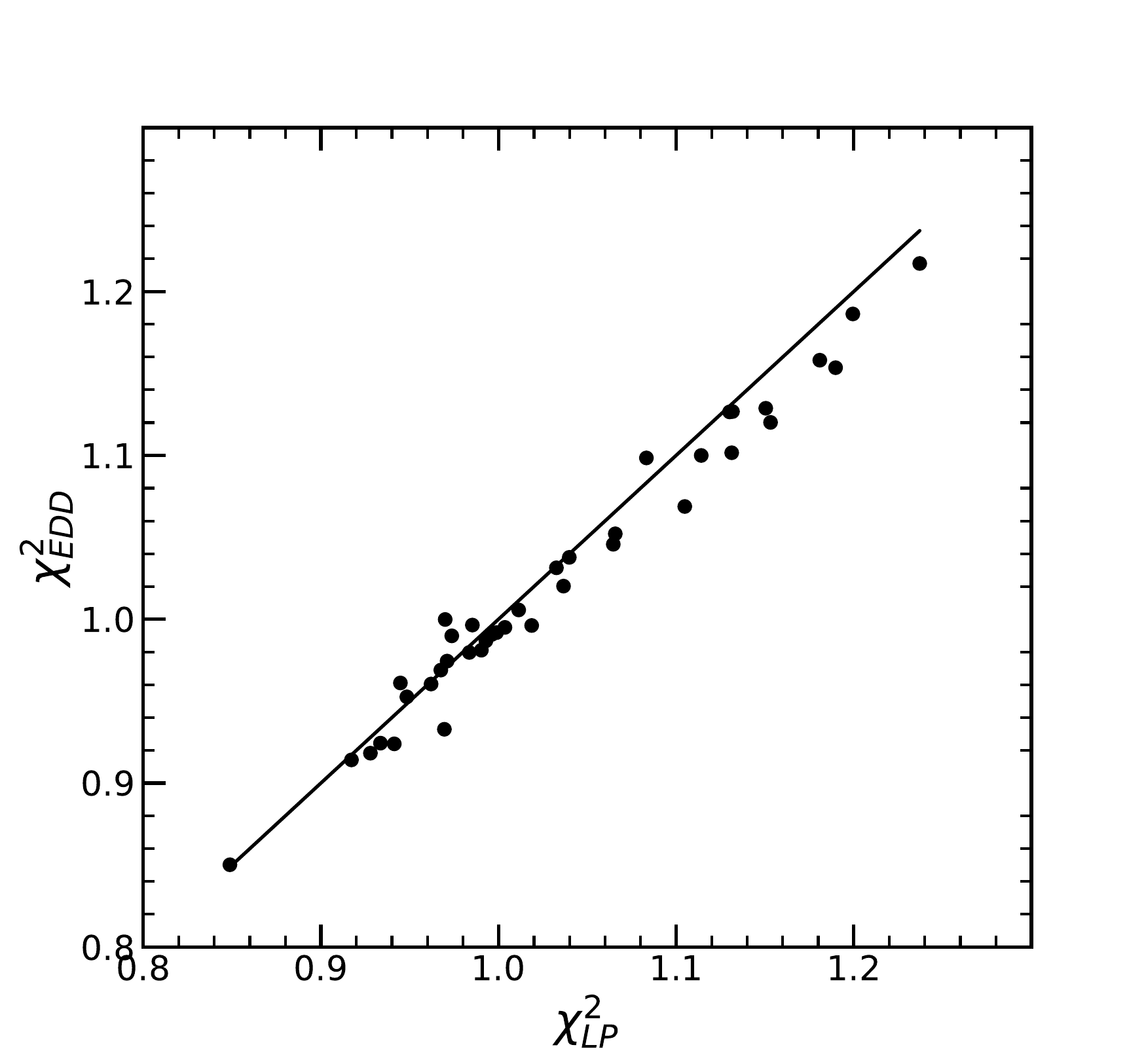}
	\includegraphics[width=0.85\columnwidth]{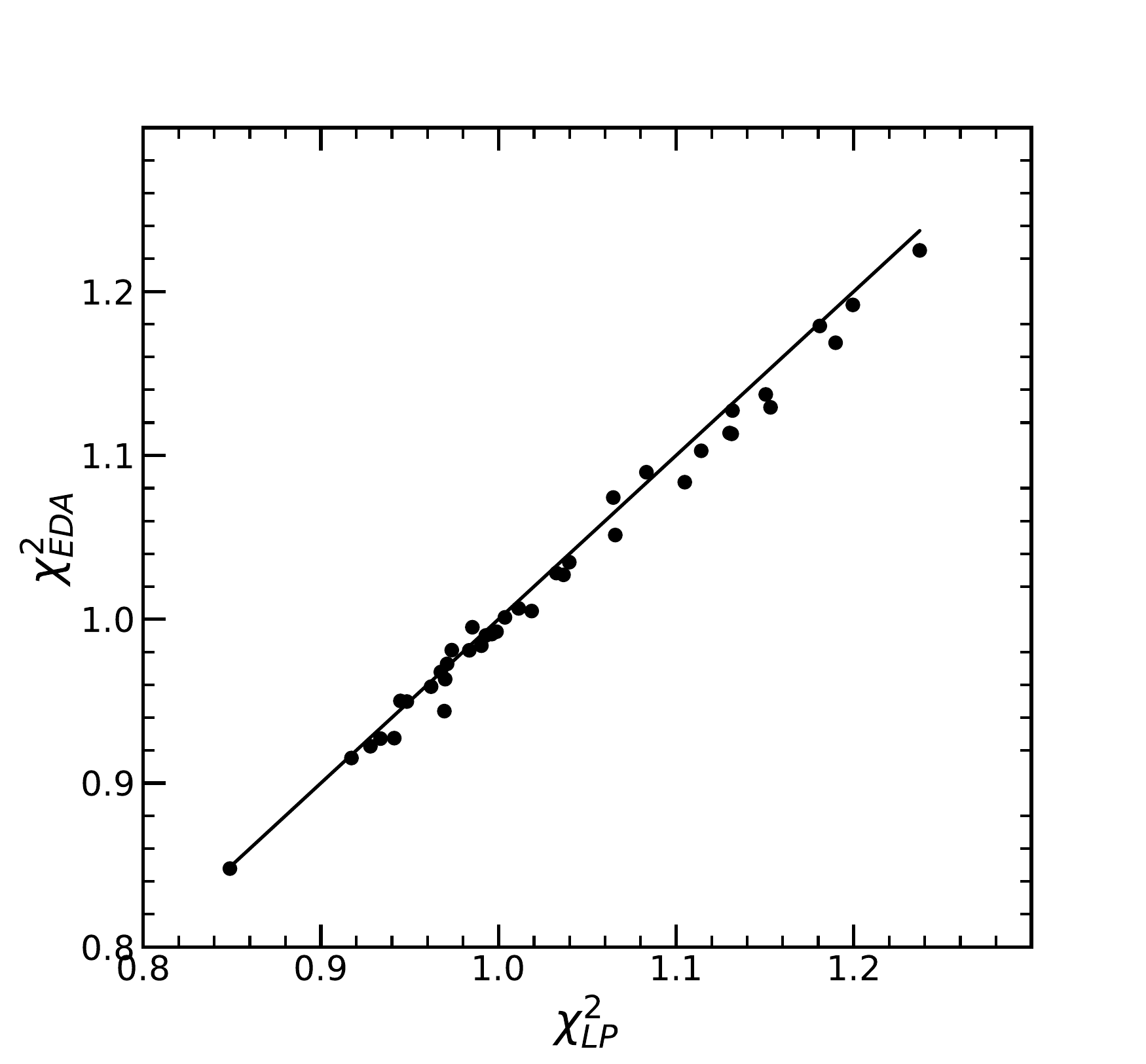}
    \caption{Comparison of reduced-$\chi^{2}$ of synchrotron convolved logparabola model with: top left panel: the reduced $\chi^2$ of $\xi_{max}$ model, top right panel: the reduced $\chi^2$ of synchrotron convolved EDD model and  bottom  left panel: the reduced $\chi^2$ of synchrotron convolved EDA model. Solid line in each panel represents identity line.}
    \label{fig:lp_gamma}
\end{figure*}

Using \autoref{eq:gmax_part}  as an input particle density, we again perform the $\chi^2$-fit of the combined SXT, LAXPC10 and LAXPC20 spectrum with \emph{$synconv \otimes n(\xi)$} model (here after $\xi_{max}$ model) in all the 39 time bins. In each time bin the fit was carried with three free parameters,  $\xi_{max}$, $p$,  and the normalization  $\mathbb{N}$ defined as
\begin{equation} \label{eq:norm_ximax}
\mathbb{N}  = \frac{\delta^3(1+z)}{d_L^2}V\,\mathbb{A} Q_0\tau_{acc}{\gamma_0^{p-1}} \mathbb{C}^{p/2}
\end{equation}
We find that this model provides an acceptable fit to all time segments, although the reduced $\chi^2$ is typically larger than that for the log-parabola model. The reduced $\chi^2$ for the two models are plotted against each other in the top left panel of \autoref{fig:lp_gamma}.

\begin{figure*}
	 	\includegraphics[height=6.95cm, width=0.95\columnwidth]{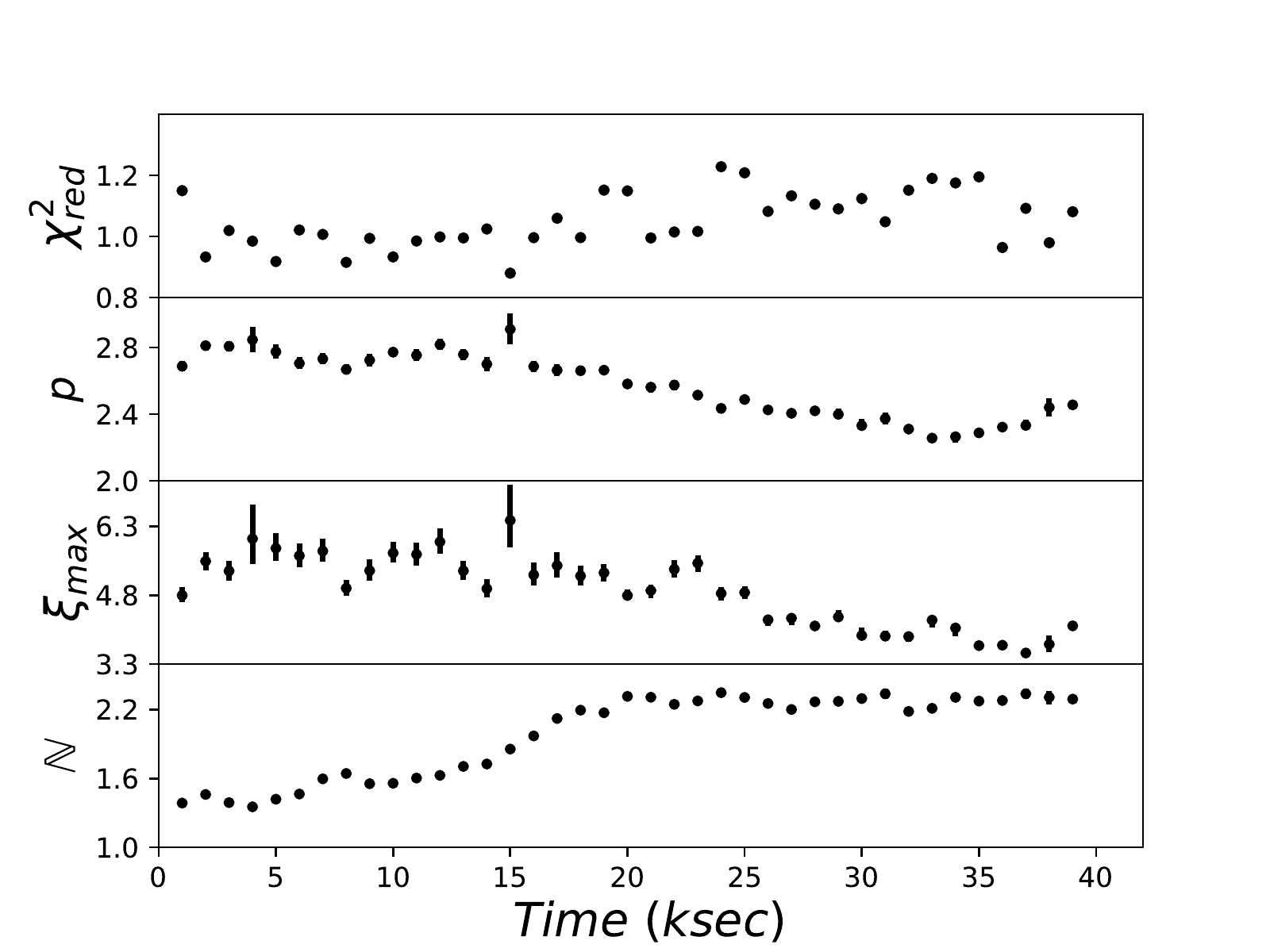}
		\includegraphics[width=0.88\columnwidth]{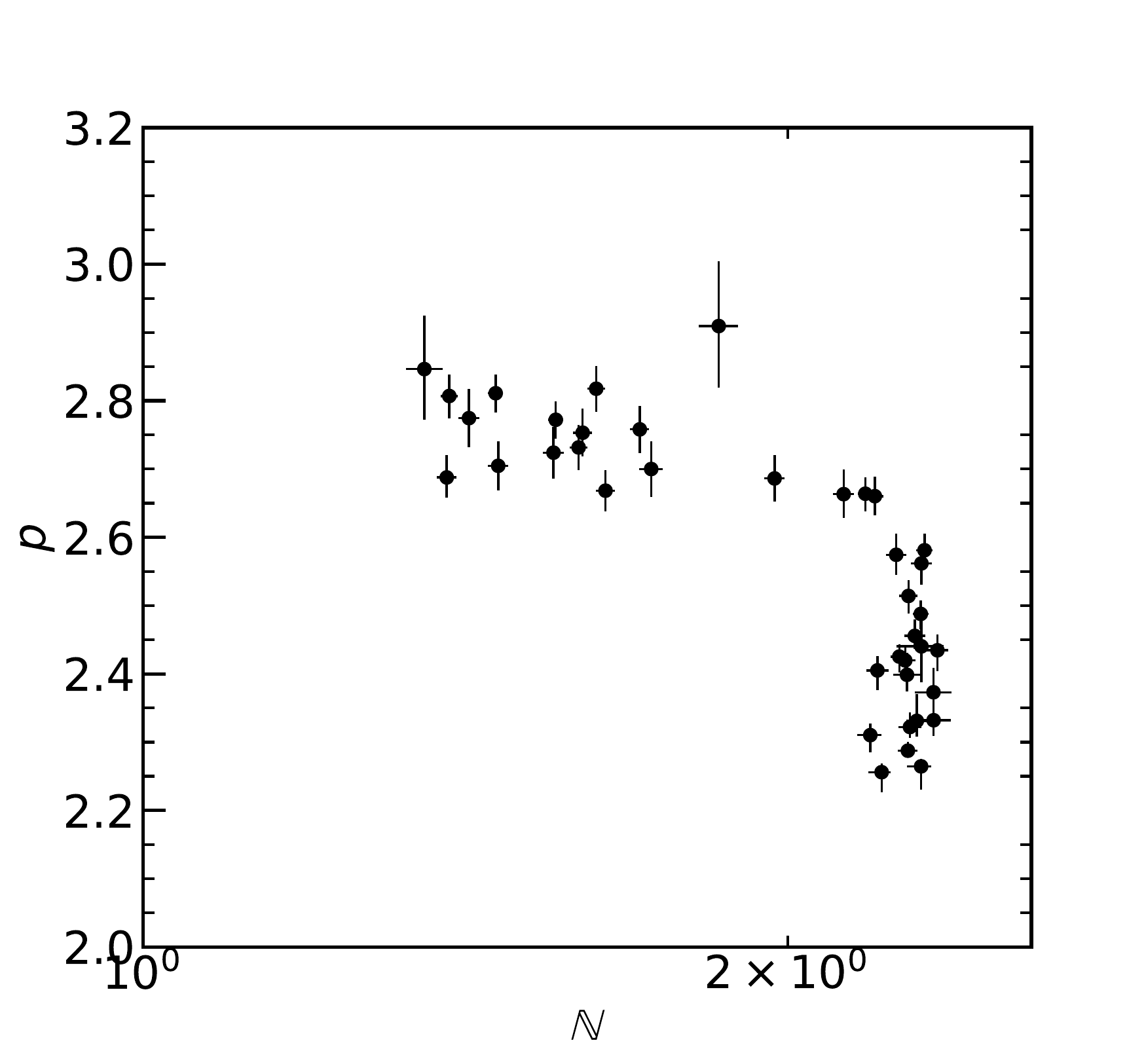}
		\includegraphics[width=0.88\columnwidth]{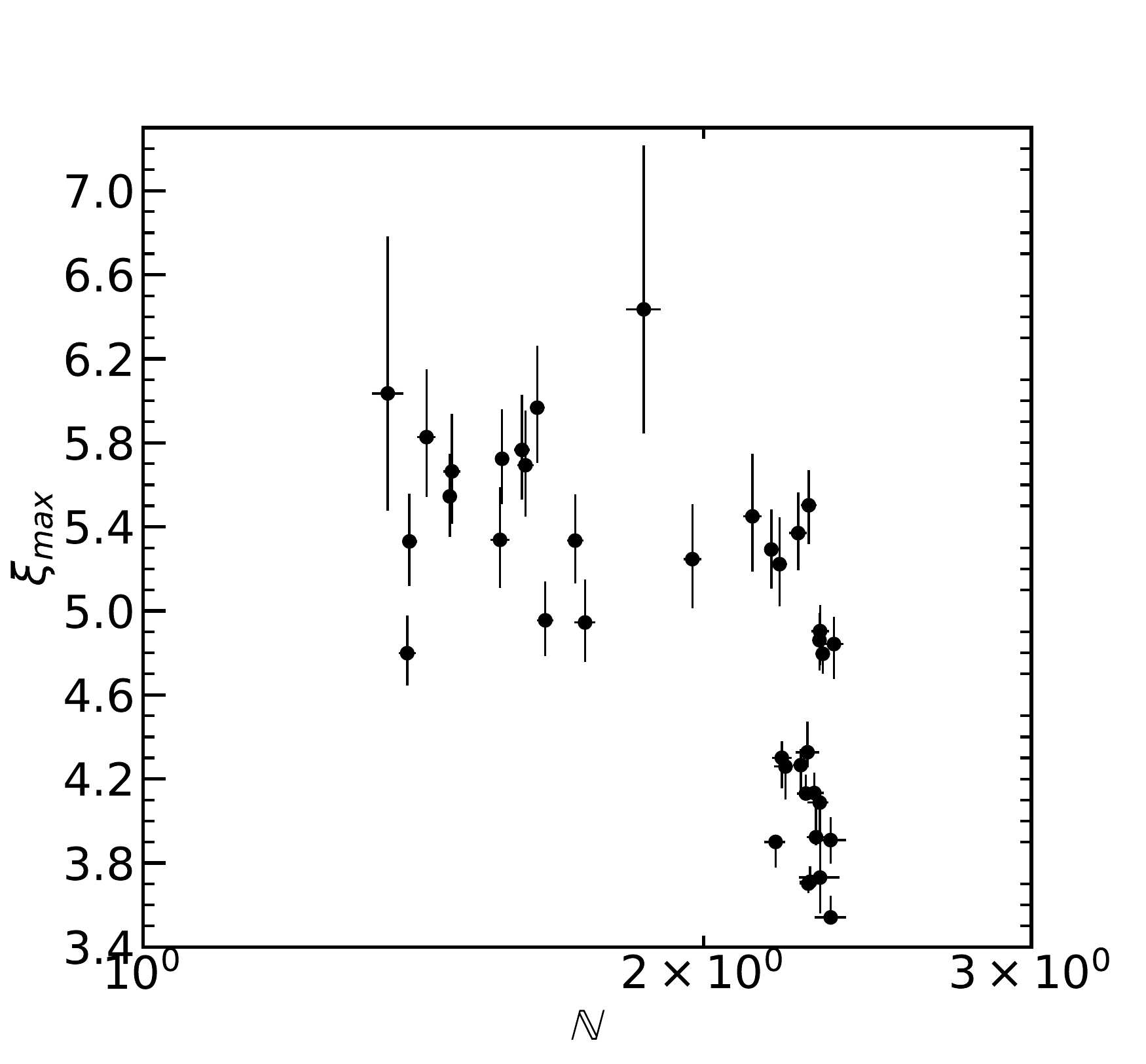}
		\includegraphics[width=0.88\columnwidth]{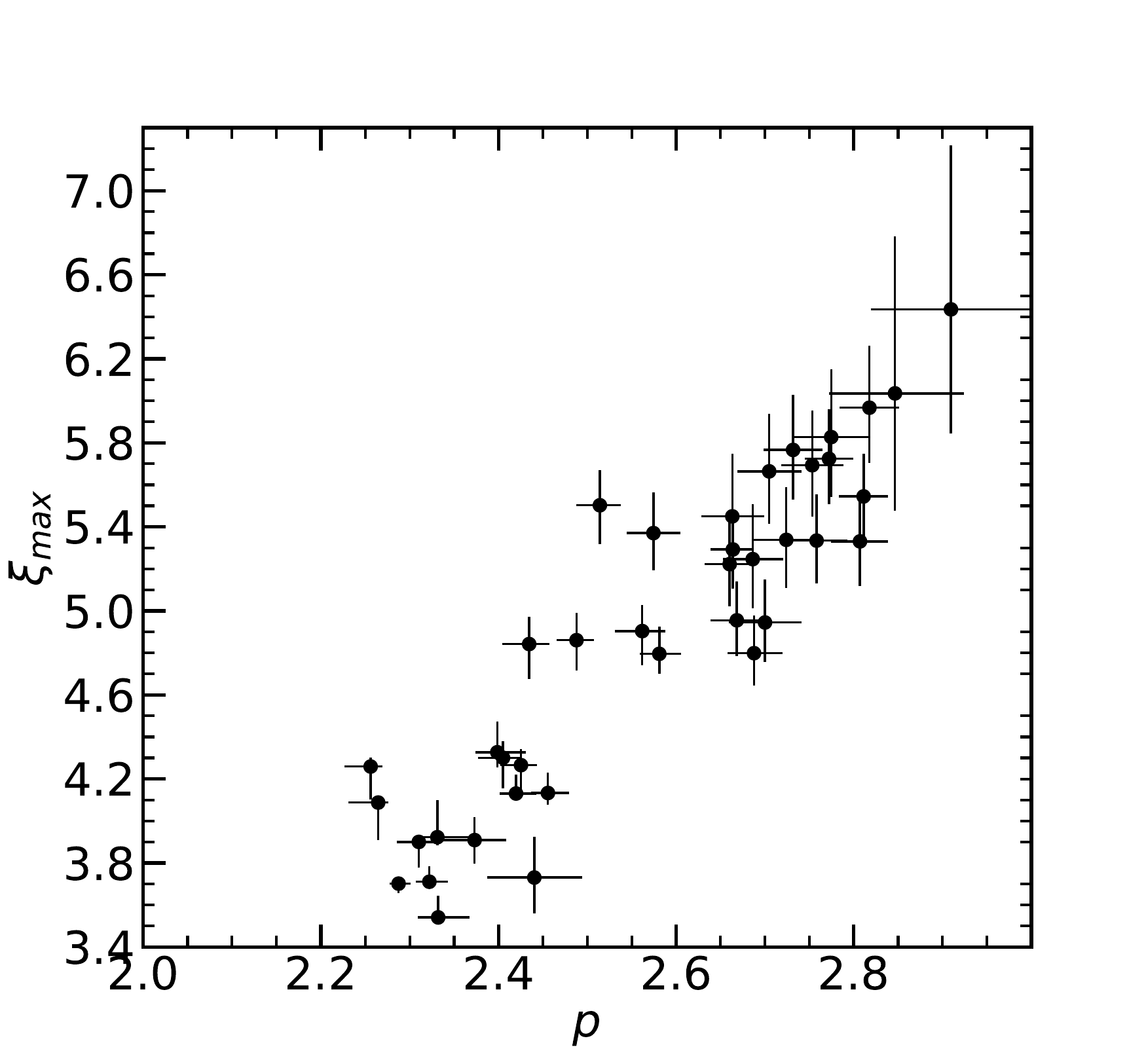}
		\includegraphics[width=0.88\columnwidth]{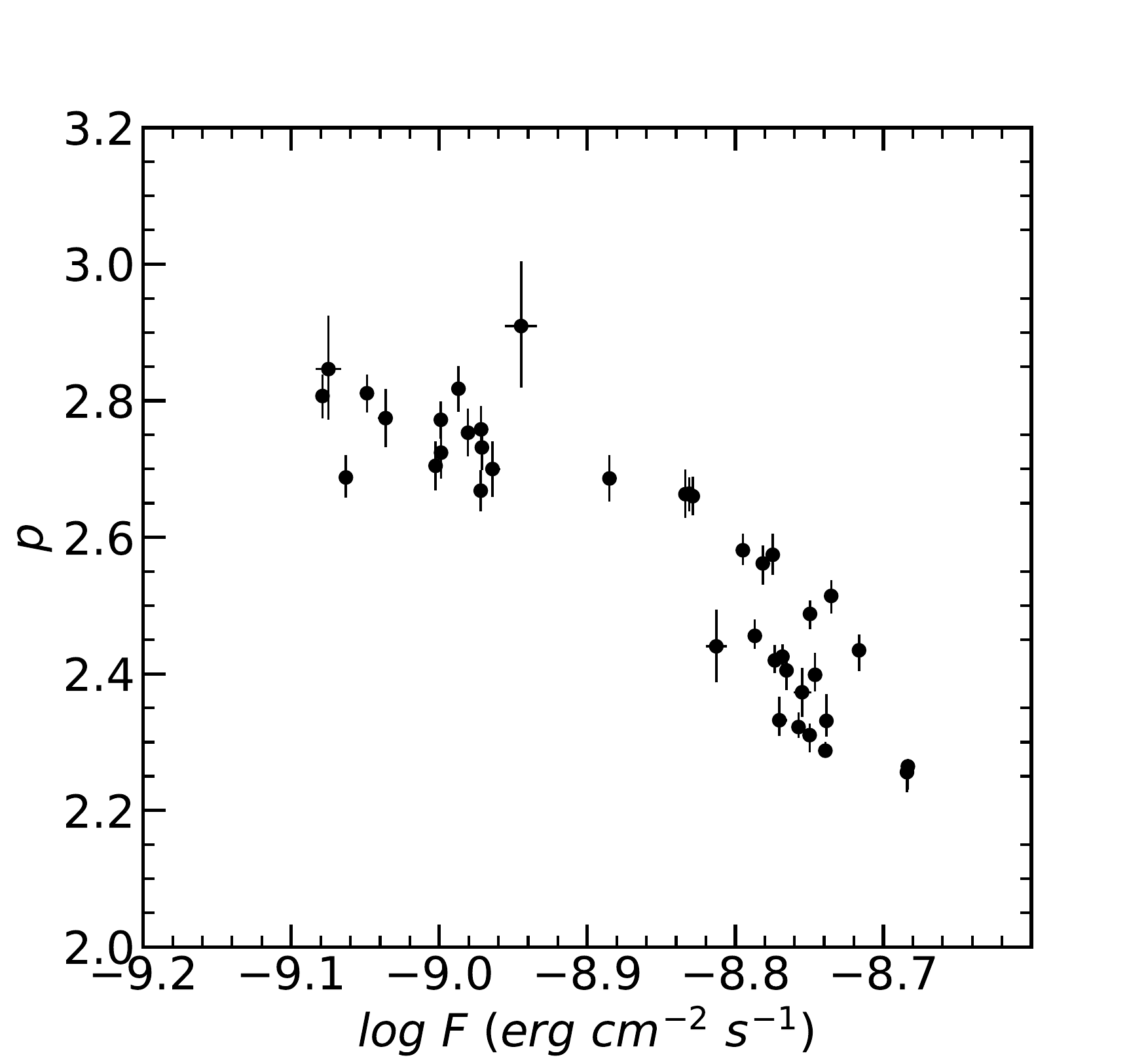}
		\includegraphics[width=0.88\columnwidth]{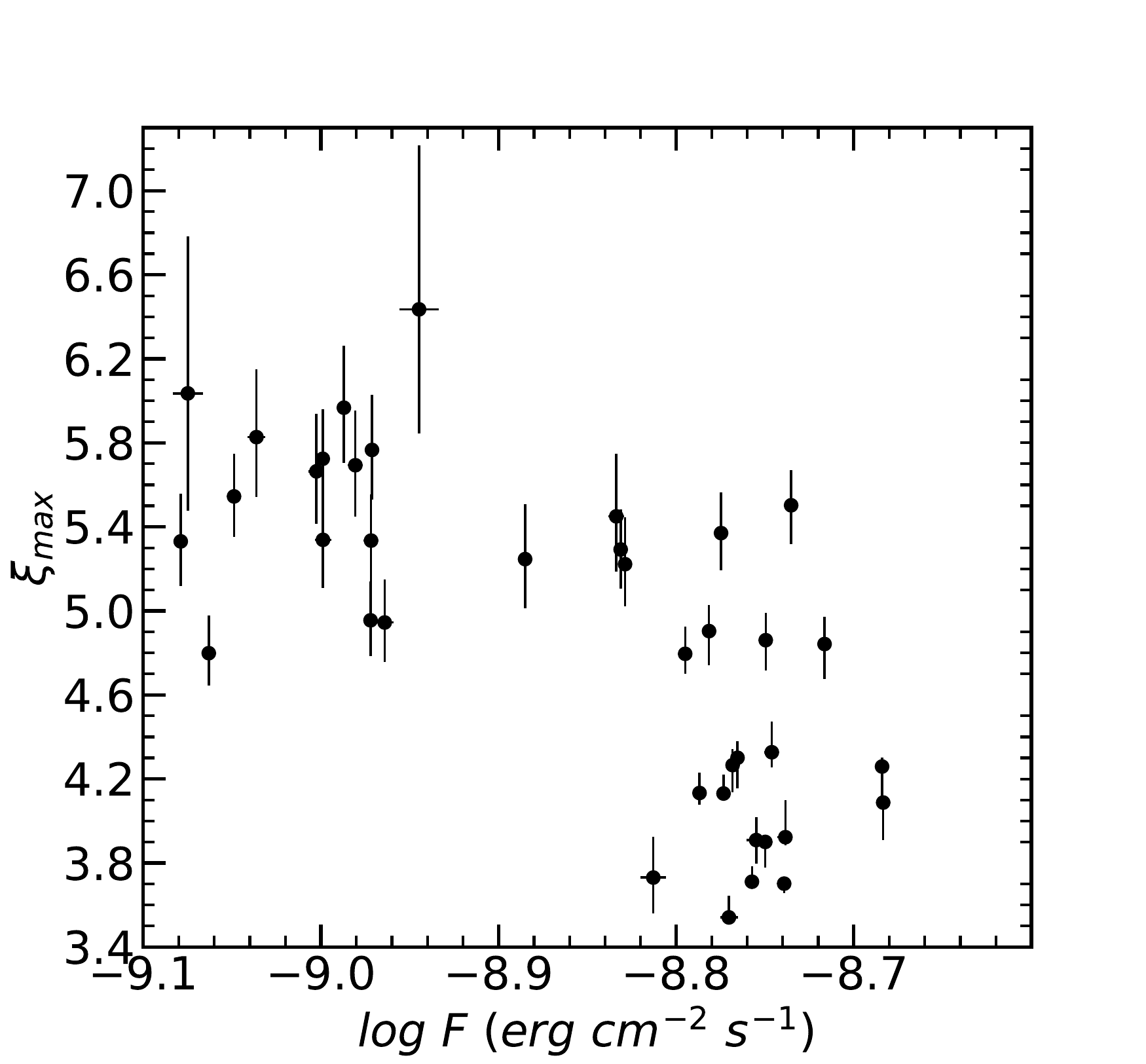}
    \caption{Top left panel: Best fit results obtained by fitting the wideband spectra in each time segment with $\xi_{max}$ model.  Top right panel: scatter plot between best fitted $\mathbb{N}$ and $p$ values,  middle  left  panel: scatter plot between best fitted $\mathbb{N}$ and $\xi_{max}$ values and middle right panel: scatter plot between best fitted  $\xi_{max}$ and $p$ values,
   bottom left panel: scatter plot between flux, $F_{0.5-18.0 \ kev}$ and best fitted $p$ values, bottom right panel: scatter plot between flux, $F_{0.5-18.0 \ kev}$ and best fitted $\xi_{max}$ values.}
    
    
    \label{fig:emax_norm}
\end{figure*}

In \autoref{Tab:loggammamax} (right), we provide the best-fit parameters for all the time segment and the reduced $\chi^{2}$ value using $\xi_{max}$ model.
A strong anti-correlation is obtained between $p$ and flux, $F_{0.5-18.0 \ kev}$ with r-value and p-value as  $\sim -0.89$  and 9.62 $\times10^{-15}$, respectively (bottom left panel of \autoref{fig:emax_norm}). Also a moderate  anti-correlation is obtained between $\xi_{max}$ and flux, $F_{0.5-18.0 \ kev}$ with r-value and p-value as $\sim -0.68$ and 1.4 $\times10^{-6}$, respectively (right bottom panel of \autoref{fig:emax_norm}).
 The correlation plots between best fitted parameters are shown in \autoref{fig:emax_norm}.
The correlation between the fitted parameters is similar to the results obtained for the log-parabola model. $\xi_{max}$ and the index $p$ are correlated to each other and show little variation with normalization till
$\mathbb{N} \sim 2$ and rapidly decrease for higher values. The tight correlation between $p$ and $\xi_{max}$ implies that variation in the index is perhaps the primary driver of the variability.

In the framework of this model, variation of the index, $p$ can only occur if $\tau_{esc}$ or $\tau_{acc}$ change. Now, there is no direct dependence of $\xi_{max}$ on $\tau_{esc}$, while $\xi_{max}$ is inversely proportional to $\tau_{acc}$ which in turn is directly proportional to $p$. Thus, variation in $\tau_{acc}$ should lead to $\xi_{max}$ being inversely proportional to $p$ which is completely contrary to the observed positive correlation between best fit values of $p$ and $\xi_{max}$ (\autoref{fig:emax_norm}). Thus, although the model provides reasonable spectral fitting to the data, the correlation between $p$ and $\xi_{max}$ can be used to rule it out. Another issue is the variation of the normalization with index, $\mathbb{N} \propto (\gamma_0^2 \mathbb{C})^{p/2}$ (Equation \ref{eq:norm_ximax}), which implies that a change in index $\Delta p$ should result in a corresponding change in normalization $\Delta \mathbb{N}/\mathbb{N} = log (\xi_0^2) \Delta p /2$, where $\xi_o^2 \equiv \gamma_0^2\mathbb{C}$. Now since $\xi_o^2$ corresponds to the observed energy of the photon (in keV) produced by electrons with energy $\gamma_0$, it should be a significantly small number since $\gamma_o$ is small compared to the $\gamma$ required to produce X-ray photons. Thus, $|log (\xi_0^2)| >> 1$ and there should be a significant variation of the normalization when the index changes. However, the spectral fit parameters show that when $p$ changes from $2.2$ to $2.6$, the normalization varies only by a few percentage (top right panel of Figure \ref{fig:emax_norm}). As pointed out by \citet{2021MNRAS.504.5485S}, the power-law solution to Equation \ref{kinetic} (for $\gamma < \gamma_{max}$), approximately pivots around $\gamma_0$ when $p$ is varied, leading to large variation in the
number density at $\gamma (>> \gamma_0$), which is typically not observed.

The model can be perhaps made compatible with the data if it is assumed that the acceleration time-scale depends on the magnetic field such that  $\tau_{acc} \propto B^{-n}$ or $B \propto (p-1)^{-1/n}$. In that case,  $\xi_{max} = \sqrt{\mathbb{C}}/(\beta_s \tau_{acc}) \propto B^{n-3/2} \propto (p-1)^{-(n-3/2)/n}$. Thus, $\xi_{max}$ would be positively correlated with $p$ if $n < 3/2$ and will be linearly correlated if $n \sim 3/4$. However, this will not address the issue of the normalization being insensitive the variations of the spectral index.

\subsubsection{Energy dependent diffusion (EDD) model}
The curvature in the spectrum can also be fitted by considering  energy dependence of the escape time-scale $\tau_{esc}(\gamma)$. The situation can be visualized by assuming that the electrons gain relativistic energy by crossing a shock front and this acceleration ceases when they diffuse away from the shock. If this diffusion is occuring within a region having a tangled magnetic field, the diffusion coefficient may depend on the gyration radius which in turn depends on the energy of the electron. Thus, the diffusion coefficient or equivalently the escape time-scale $\tau_{esc}$ may be energy dependent. We can parameterize this energy dependence by assuming that $\tau_{esc} = \tau_{esc,R} \left(\frac{\gamma}{\gamma_R}\right)^{-\kappa} $. Note that $\tau_{esc}$ cannot be larger than the free-streaming value and this limit will be reached when the gyration radius is of the order of size of the region. We can then identify $\gamma_R$ as the energy at which this limit is reached, $\tau_{esc,R}$ as the free streaming value and restrict the validity of the dependence to $\gamma < \gamma_R$. The solution to Equation \ref{kinetic}, with this dependence of the  escape time-scale and neglecting the synchrotron energy loss, becomes
\begin{equation}
  n(\xi)=Q_o \tau_{acc}\sqrt{\mathbb{C}} \xi^{-1}\rm{exp}\left[-\frac{\eta_R}{\kappa}\,\left(\left(\frac{\xi}{\xi_R}\right)^{\kappa}-\left(\frac{\xi_0}{\xi_R}\right)^{\kappa}\right)\right]
  \label{n_EDDt}
\end{equation}
where $\xi_R = \sqrt{\mathbb{C}}\gamma_R$, $\xi_0 = \sqrt{\mathbb{C}}\gamma_0$ and $\eta_R \equiv \tau_{acc}/\tau_{esc,R}$. For small value of $\kappa <<1$, the solution reduces to a log-parabola form
\begin{equation}
  n(\xi) \propto (\xi/\xi_R)^{-\eta_R-1 -\eta_R\kappa log(\xi/\xi_R)}
\end{equation}
and for $\xi \rightarrow 0$, the distribution is identical to that given in Equation \ref{eq:gmax_part} for $\gamma << \gamma_{max}$ and $p = \eta_R+1$ i.e., when the escape time-scale is energy independent and equal to the free-streaming value of $\tau_{esc,R}$. The particle energy distribution given in Equation \ref{n_EDDt} depends on a number of parameters which would be degenerate when it is used for spectral fitting. Hence, we recast the distribution as 
\begin{equation}
\label{eq:eddcr1}
	n(\xi)=K^\prime \xi^{-1}\rm{exp}\left[-\frac{\psi}{\kappa}\,\xi^{\kappa}\right]
\end{equation}
where  
$\psi=\eta_R\left(\mathbb{C}\gamma_R^2\right)^{-\kappa/2} = \eta_R \xi_R^{-\kappa} $,   and the normalization,
\begin{equation}
K^\prime = Q_0 \tau_{acc} \sqrt{\mathbb{C}} \rm{exp}\left[\frac{\eta_R}{\kappa}\,\left(\frac{\gamma_0}{\gamma_R}\right)^{\kappa}\right]
\end{equation}

We performed a $\chi^2$-fit to the combined SXT,
LAXPC10 and LAXPC20 spectra using \emph{$synconv \otimes n(\xi)$} model with input particle density given by  \autoref{eq:eddcr1} and we refer to  this model as synchrotron convolved EDD model. The free parameters for the model are $\psi$, $\kappa$, and $\mathbb{N}$ given by 
\begin{equation}\label{eq:norm_eed}
   \mathbb{N}  = \frac{\delta^3(1+z)}{d_L^2} V \mathbb{A}K^\prime
\end{equation}

\begin{figure*}
	\includegraphics[width=0.9\columnwidth]{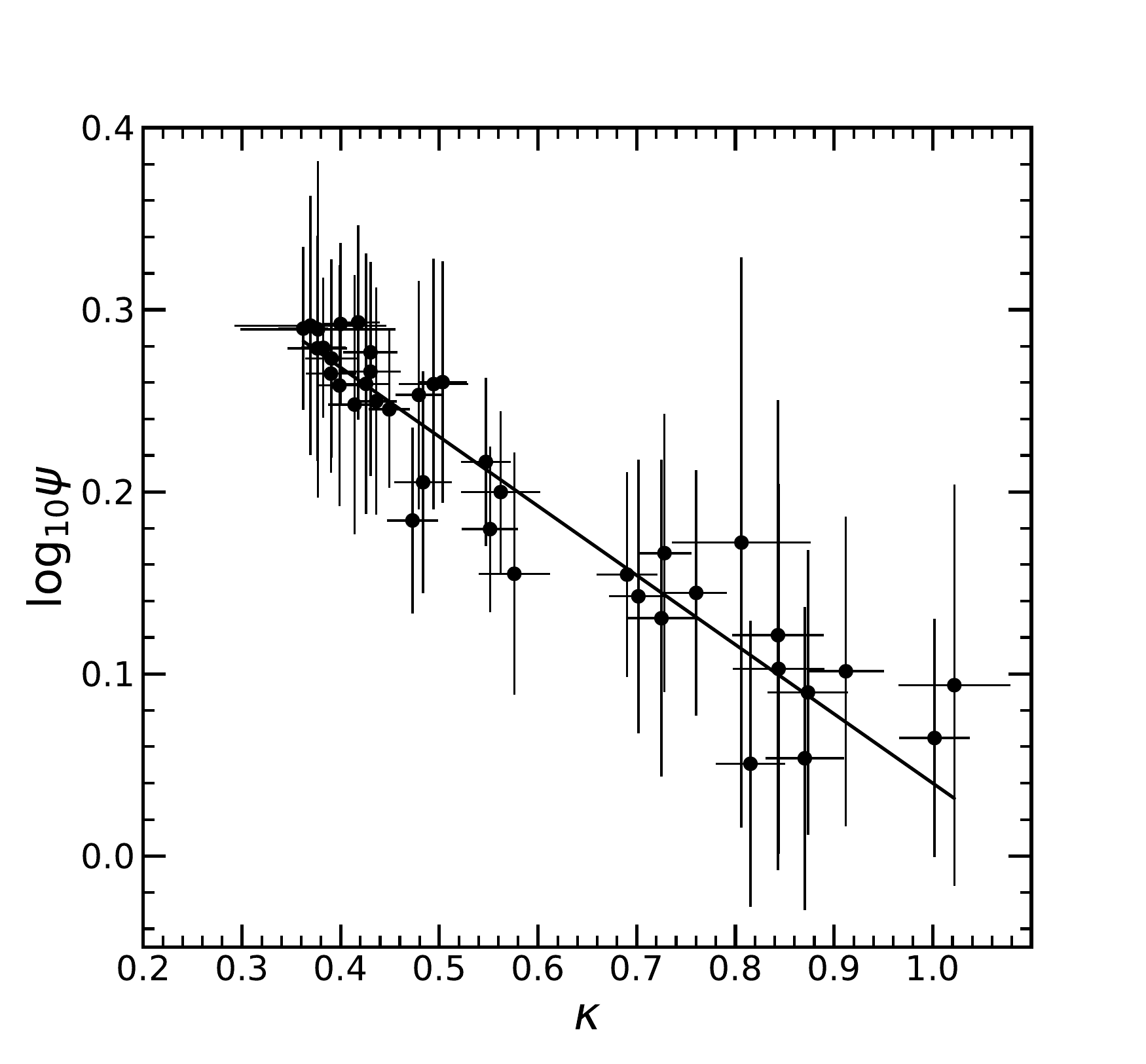}
	\includegraphics[width=0.9\columnwidth]{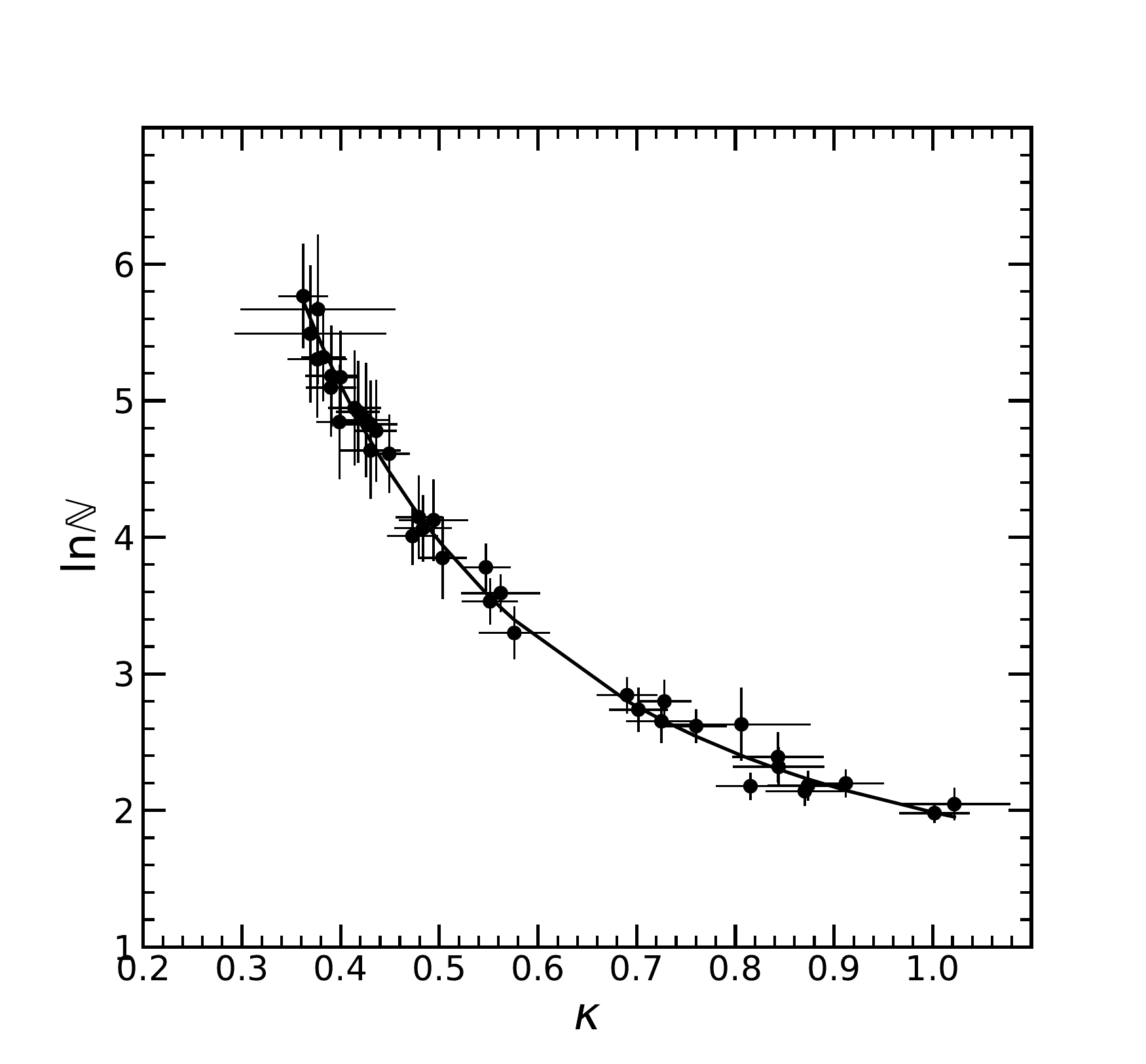}
	\includegraphics[width=0.9\columnwidth]{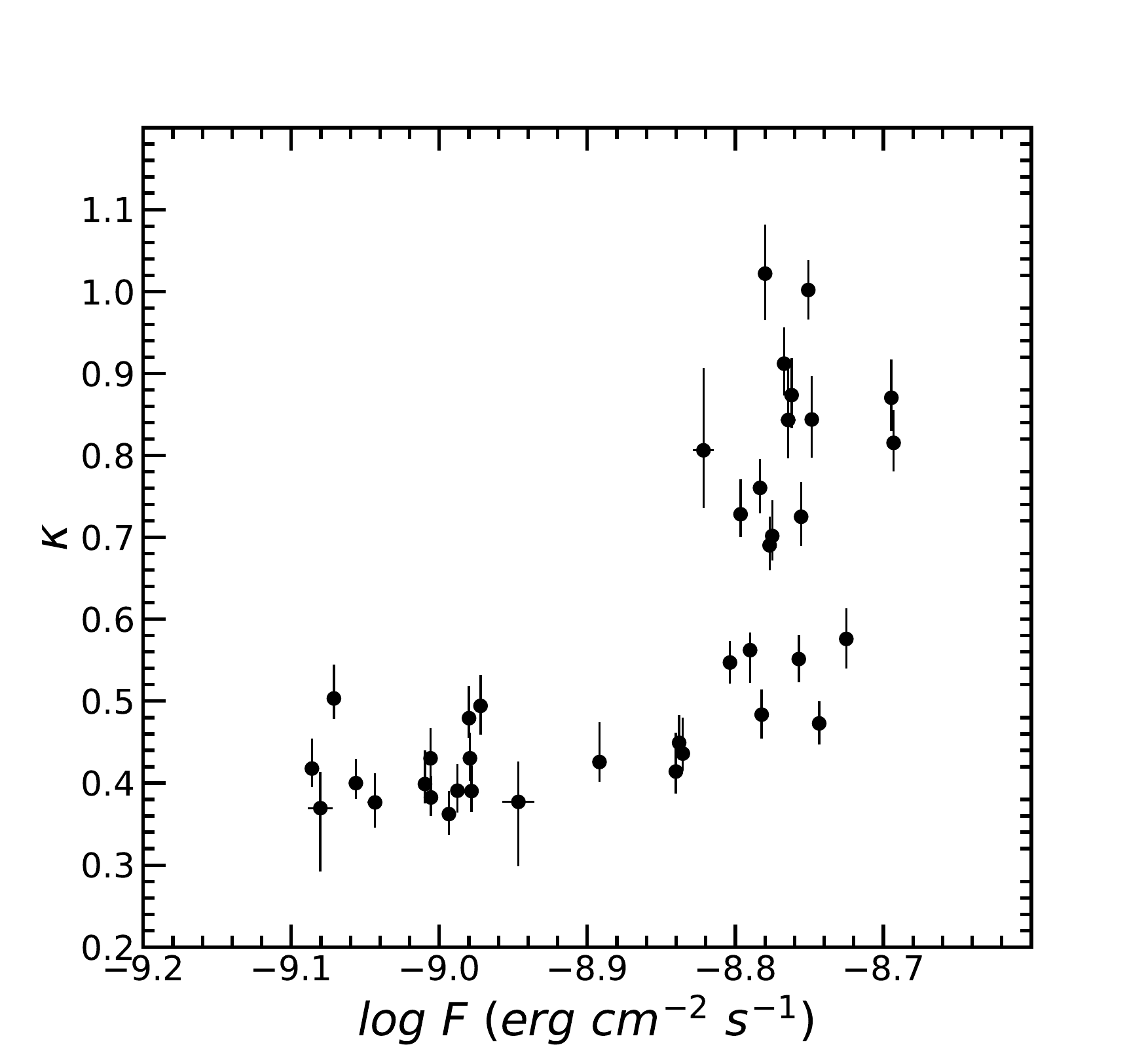}
	\includegraphics[width=0.9\columnwidth]{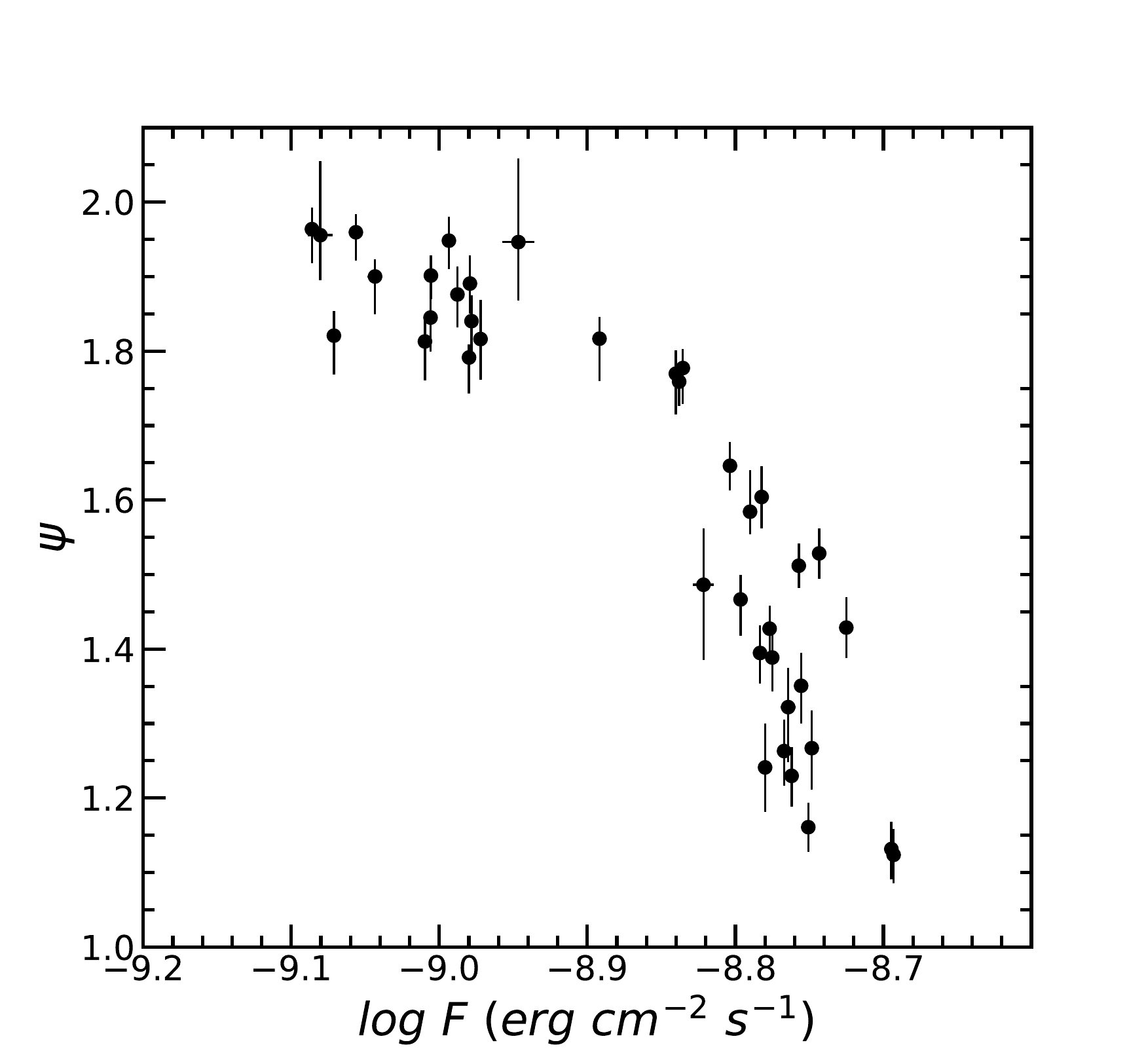}

\vspace{0.5cm}
    \caption{Best fit results obtained by fitting the wideband spectra in each time segment with synchrotron convolved EDD model. Top left panel : Plot between best fitted $\kappa$ and $\log_{10}\psi$ values, Top right panel: plot between  $\ln \mathbb{N}$ and $\kappa$ values. Solid curve in each panel is best fitted function as described in the text.
     Bottom left panel: scatter plot between flux, $F_{0.5-18.0 \ kev}$ and best fitted $\kappa$ values, bottom right panel: scatter plot between flux, $F_{0.5-18.0 \ kev}$ and best fitted $\psi$ values.}

    \label{fig:Eddxi_kappa_norm}
\end{figure*}

The reduced-$\chi^2$ obtained in all the time-segments suggests that the synchrotron convolved EDD model provides good fit to the broadband X-ray spectrum. For comparison with the logparabola fit, we have plotted the reduced-$\chi^{2}$ values obtained from the synchrotron convoluted log-parabola against the reduced-$\chi^2$ of synchrotron convolved EDD model in top right panel of  \autoref{fig:lp_gamma}, the figure suggest that both log-parabola and EDD model fits the X-ray spectrum well,  while the EDD model provides comparatively better fit to the spectra  of some of the  time segments.
In \autoref{Tab:edaedd} (right), we provide the best-fit parameters for all the time segment and the reduced $\chi^{2}$ values using EDD model.
A strong anti-correlation is obtained between $\psi$ and flux, $F_{0.5-18.0 \ kev}$ with r-value and p-value as  $\sim  -0.90$ and 3.9 $\times10^{-15}$, respectively (bottom left panel of \autoref{fig:Eddxi_kappa_norm}). On the other hand a strong correlation is obtained between $\kappa$ and flux, $F_{0.5-18.0 \ kev}$ with r-value and p-value as $\sim 0.84$ and 1.5 $\times10^{-11}$, respectively (bottom right panel of \autoref{fig:Eddxi_kappa_norm}). 
Since $\psi = \eta_R \xi_R^{-\kappa} $ one expects log$\psi$ to be linearly and inversely proportional to $\kappa$ and this is indeed the case. The left panel of Figure \ref{fig:Eddxi_kappa_norm} shows the variation of log$_{10}\psi$ with $\kappa$ and the straight line fit, log$_{10}\psi = -0.38 \kappa + 0.42$, which implies that $\eta_R \sim 2.6$ and $\xi_R \sim 2.4$. Moreover the normalization as given by Equation \ref{eq:norm_eed} implies that log$\mathbb{N} = \left[\frac{\eta_R}{\kappa}\,A^{\kappa}\right] + B$. This form does represent the variation of the normalization as shown in the right panel of Figure \ref{fig:Eddxi_kappa_norm}, where the solid line represents the form with $A = 0.26$ and $B  = 1.3$ and $\eta_R = 2.6$. This provides an estimate of $\gamma_0 \sim 0.26 \gamma_R$.
While correlations between the best fit spectral parameters are consistent with the predictions of the model, the inferred physical values are not quite as expected. The model is not consistent for electrons with energy $> \gamma_R$. However, the observed photon energy corresponding to $\gamma_R$  is estimated to be $\xi_R^2 \sim 5.75$ keV, which is within the energy range considered for spectral fitting. Moreover the injection energy $\gamma_0$ is not significantly smaller than $\gamma_R$. These issues will be discussed further in the last section.

\subsubsection{Energy dependent acceleration model}
\begin{figure*}
		\includegraphics[width=0.9\columnwidth]{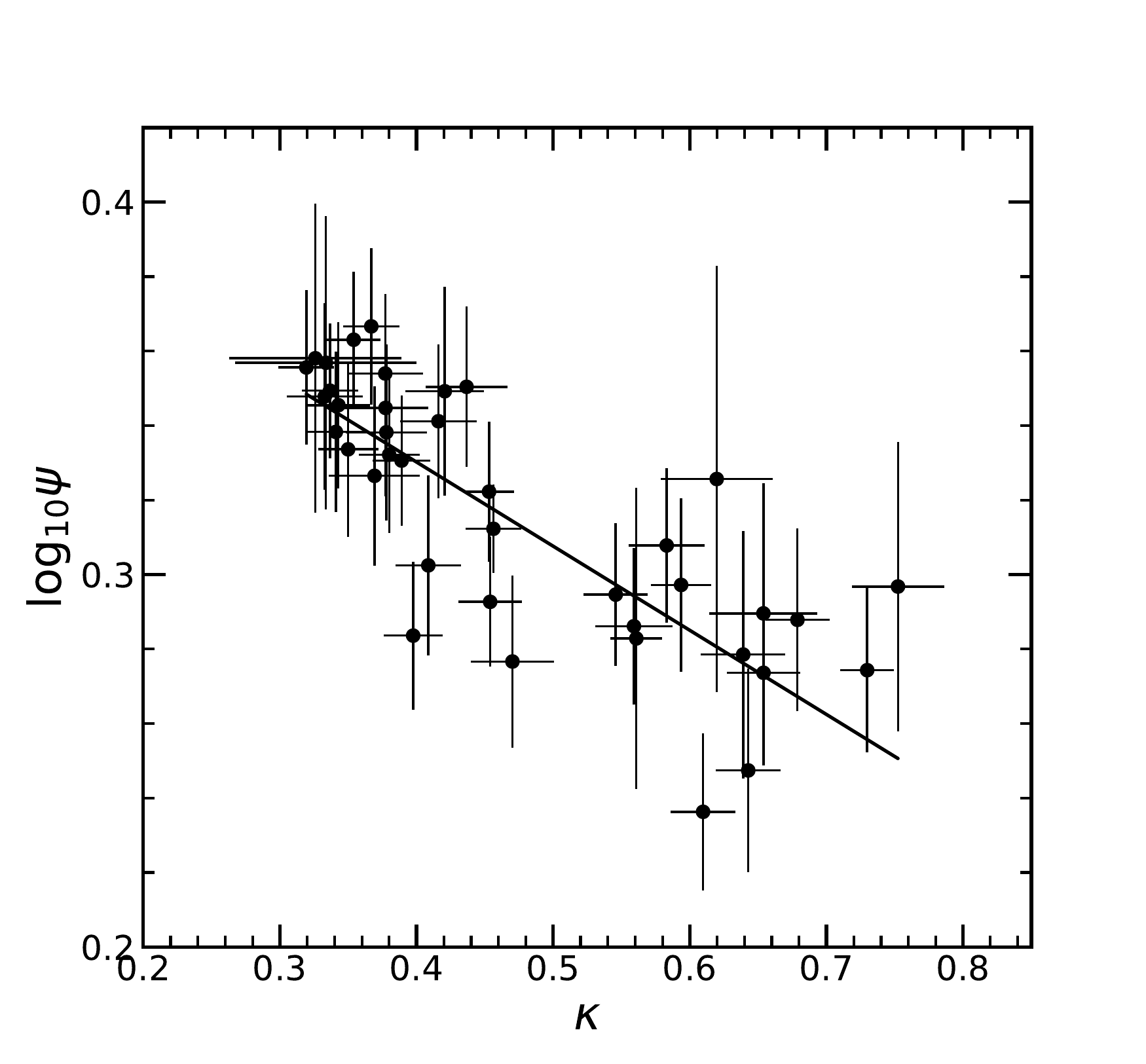}
	\includegraphics[width=0.9\columnwidth]{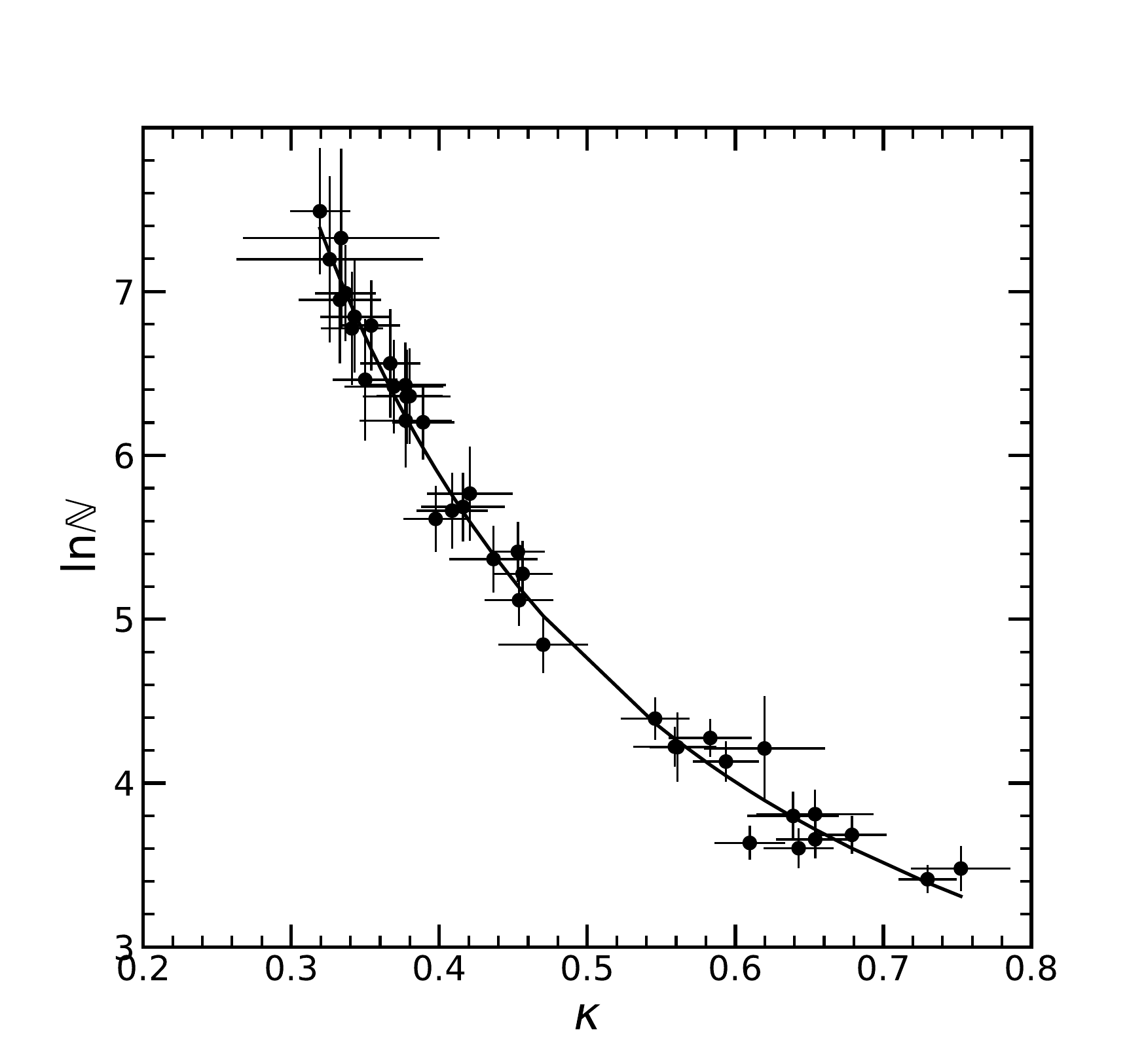}
	\includegraphics[width=0.9\columnwidth]{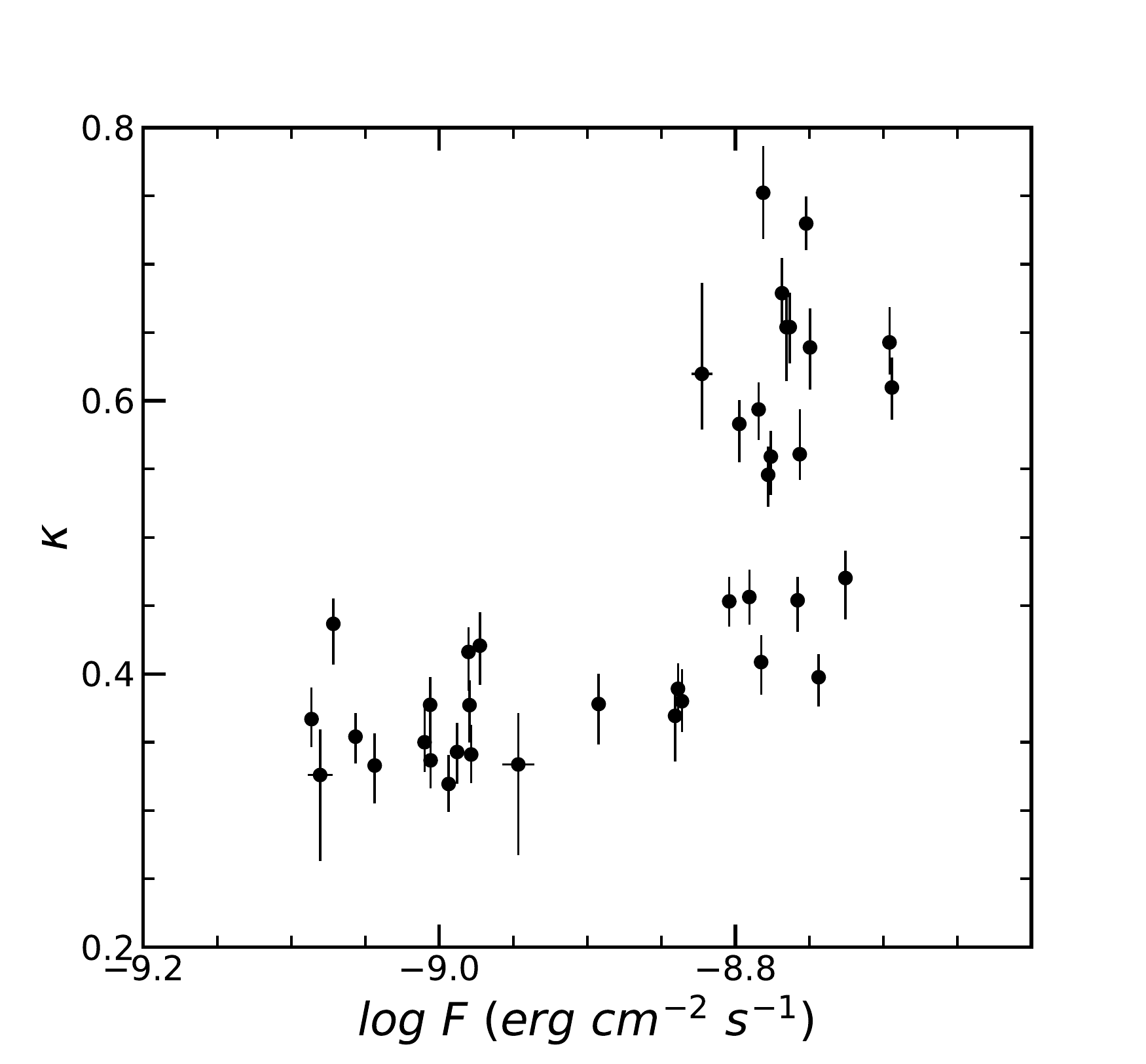}
	\includegraphics[width=0.9\columnwidth]{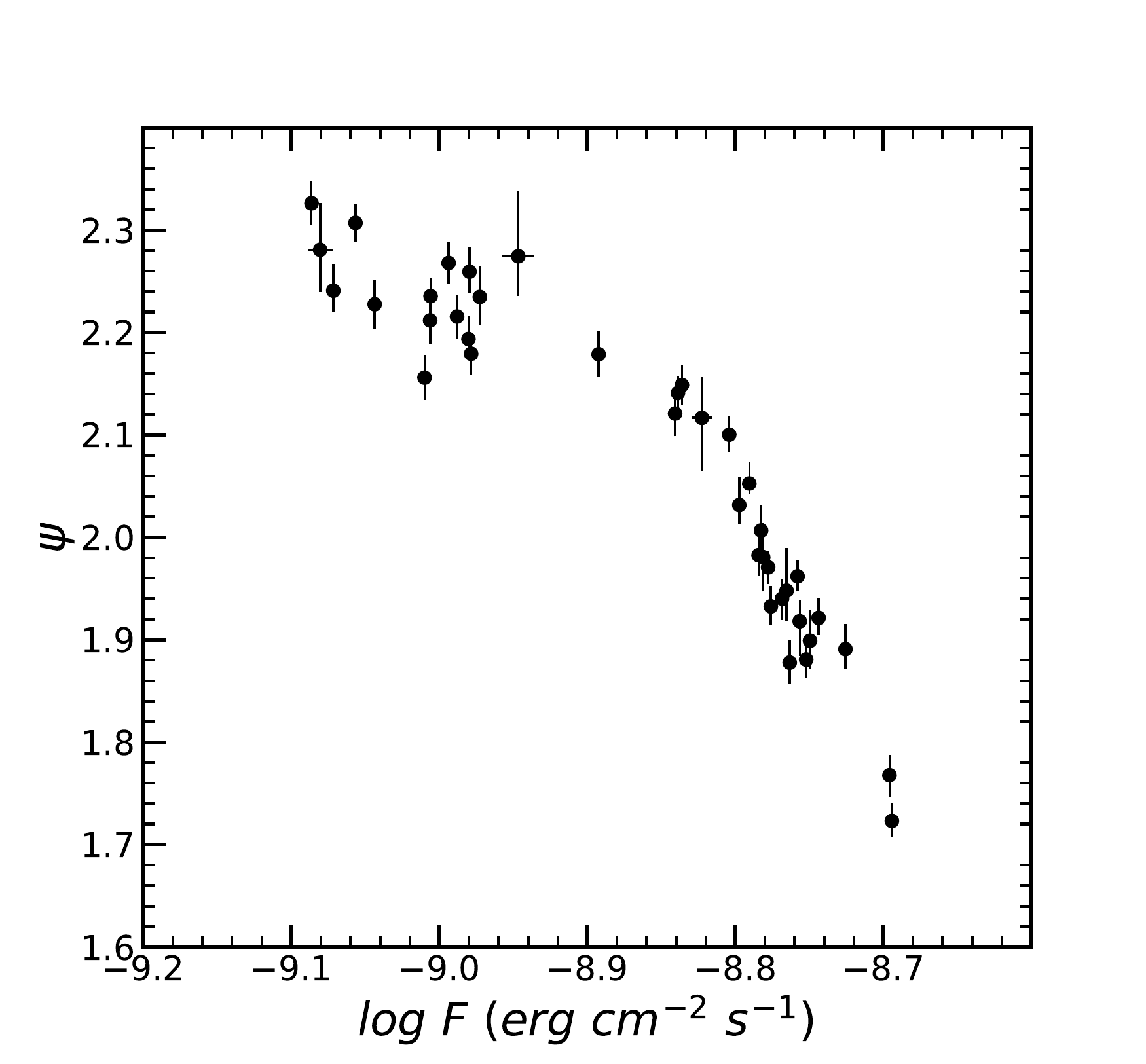}
    \vspace{0.5cm}
    \caption{Best fit results obtained by fitting the wideband spectra in each time segment with synchrotron convolved EDA model. left panel : Plot between best fitted $\kappa$ and $\log_{10}\psi$ values,  right panel:  plot between  $\ln \mathbb{N}$ and $\kappa$ values. Solid curve in each panel is best fitted function as described in the text.
    Bottom left panel: scatter plot between best fitted average flux in the range 0.5 -- 18.0 keV and $\kappa$ values, bottom right panel: scatter plot between best fitted average flux in the range 0.5 -- 18.0 keV and $\psi$ values.}
    \label{fig:edaxi_kappa}
\end{figure*}

We next consider a scenario in which the acceleration process is energy dependent \citep{2004Massaro,2006A&A...448..861M,2007Tramacere}. While \citet{2004Massaro} considered that the probability that an electron accelerates, decreases with energy, here we invoke an energy dependent acceleration time-scale i.e. $\tau_{acc} = \tau_{acc,R}\left(\frac{\gamma}{\gamma_R}\right)^\kappa$. If we assume that the increase in the time-scale with energy is related to the increasing gyration radius of the particle, then we have to identify $\gamma_R$ as the energy corresponding to a gyration radius equal to the size of the system i.e., $\gamma_R$ would have the same meaning as it had for the energy dependent diffusion case. 
If we neglect the synchrotron energy loss, then the steady state solution of \autoref{kinetic} for such an energy dependent $\tau_{acc}$  becomes
\begin{equation}\label{eq:edatxi}
n(\xi)=Q_0\tau_{acc,R}\sqrt{\mathbb{C}}\xi_R^{-\kappa}\xi^{\kappa-1}\exp \left[-\frac{\eta_R}{\kappa}\left(\left(\frac{\xi}{\xi_R}\right)^\kappa-\left(\frac{\xi_0}{\xi_R}\right)^\kappa\right)\right]
\end{equation}
where $\xi_0 = \sqrt{\mathbb{C}}\gamma_0$, $\xi_R = \sqrt{\mathbb{C}}\gamma_R$ and $\eta_R \equiv \tau_{acc,R}/\tau_{esc}$. For small value of $\kappa <<1$, the solution again reduces to a log-parabola form
\begin{equation}
  n(\xi) \propto (\xi/\xi_R)^{-\eta_R+\kappa-1 -\eta_R\kappa log(\xi/\xi_R)} ,
\end{equation}
 and for $\kappa \rightarrow 0$ and $\gamma << \gamma_{max}$,  the solution is identical to that obtained in Equation \ref{eq:gmax_part}. Again, in order to remove the degeneracy in the parameters of the particle energy distribution given in  \autoref{eq:edatxi}, we recast the distribution as 
\begin{equation}
\label{eq:eddcr}
	n(\xi)=K^\prime \xi^{\kappa-1}\rm{exp}\left[-\frac{\psi}{\kappa}\,\xi^{\kappa}\right]
\end{equation}
where  
$\psi=\eta_R\left(\mathbb{C}\gamma_R^2\right)^{-\kappa/2} = \eta_R \xi_R^{-\kappa} $,   and the normalization,
\begin{equation}
K^\prime = Q_0 \tau_{acc,R}\sqrt{\mathbb{C}}\xi_R^{-\kappa} \rm{exp}\left[\frac{\eta_R}{\kappa}\,\left(\frac{\xi_0}{\xi_R}\right)^{\kappa}\right]
\end{equation}
We  refer to the  \emph{$synconv \otimes n(\xi)$} model  as the synchrotron convolved EDA model.
While fitting the broadband X-ray spectrum in each time bin with the synchrotron convolved EDA model, 
the parameters $\kappa$, $\psi$ and the normalization $\mathbb{N}$ were kept as free parameters. Here $\mathbb{N}$ is defined as
\begin{equation}\label{eq:norm_eda}
\mathbb{N}  = \frac{\delta^3(1+z)}{d_L^2} V \,\mathbb{A}K^\prime
\end{equation}
The synchrotron convolved EDA model also provided slightly better fit to the broadband X-ray spectra as compared to the synchrotron convolved logparabola model for some of the time segments (see bottom panel  \autoref{fig:lp_gamma}). 
In \autoref{Tab:edaedd} (left), we provide the best-fit parameters for all the time segment and the reduced $\chi^{2}$ values using EDA model.
A strong anti-correlation is obtained between $\psi$ and flux, $F_{0.5-18.0\ keV}$ with r-value and p-value as  $\sim$ - 0.95 and 1.3 $\times10^{-20}$, respectively (bottom left panel of \autoref{fig:edaxi_kappa}). On the other hand a correlation is obtained between $\kappa$ and flux, $F_{0.5-18.0 \ kev}$ with r-value and p-value as $\sim 0.78$ and 3.96$\times10^{-9}$, respectively (bottom right panel of \autoref{fig:edaxi_kappa}). 

The relation $\psi = \eta_R \xi_R^{-\kappa} $ shows that log$\psi$ should be linearly and inversely proportional to $\kappa$. The left panel of Figure \ref{fig:edaxi_kappa} shows that this is indeed satisfied although the relation has significantly larger scatter than the equivalent relation for the EDD model (\autoref{fig:Eddxi_kappa_norm}). The Figure shows the variation of log$\psi$ with $\kappa$ and the straight line fit, $\log_{10}\psi = -0.23 \kappa + 0.42$, which implies that $\eta_R \sim 2.63$ and $\xi_R \sim 1.70$. Moreover the variation of the normalization with $\kappa$ as shown in the right panel of \autoref{fig:edaxi_kappa} is in agreement with the 
 normalization as given by \autoref{eq:norm_eda},  log$\mathbb{N} = \frac{\eta_R}{\kappa}\,A^{\kappa}- \kappa\log\xi_R+B$. 
 In the Figure, the solid line represents the form with $A = 0.19$, $B  = 2.70$, $\xi_R= 1.70$, and $\eta_R = 2.63$ . 
 This provides an estimate of $\gamma_0 \sim 0.19 \gamma_R$. Thus the results obtained for the EDA model is qualitatively similar to those obtained for the EDD one.
 

\section{Summary and Discussion}
\label{sec:discus}

We have carried out time-resolved X-ray (0.5 -- 18 keV) spectral study of {\em{AstroSat}} observation of Mkn\,421 during 3 -- 8 January, 2017, when the source exhibited  large variations. The long 409 ksec observation was divided into segments of length 10 ksec and each segment was fitted by a synchrotron spectrum arising from a non-thermal distribution of electrons. Each spectra is well described by emission from a log-parabola distribution of electrons having  variable non-zero curvatures.

We investigated the possibility that the curvature in the particle distribution is due to synchrotron cooling, which causes a deviation from a power-law form at high energies and has a maximum value of Lorentz factor $\gamma_{max}$ and found that each of the spectra can be reasonably represented by the model. The index of the particle distribution at low energies was found to have a positive  correlation with the maximum energy of the particles. However, in the simplest picture, where the observed variation is primarily due to changes in the acceleration time-scale, the  correlation should have been negative. A more complex model where the magnetic field is correlated with the acceleration time-scale may produce the positive correlation observed, but fails to explain the variation of normalization with index.

We considered the possibility that the spectral curvature is because the electron escape or diffusion time-scale is energy dependent i.e., $\tau_{esc}= \tau_{esc,R}\left(\frac{\gamma}{\gamma_R}\right)^{-\kappa}$ and again found that each spectra can be well represented by this  Energy Dependent Diffusion (EDD) model. Moreover, two of the three  parameters of the model, the normalization and $\psi$ ($\psi$ being a combination of other physical parameters),  were found to be correlated with the third parameter $\kappa$ in complete accordance with the predictions of the model. The matching of the expected to the observed correlations allow for an estimation of the characteristic synchrotron photon energy $\xi_{R}^2 \sim 5.75$ keV arising from electron with energy $\gamma_R$ and that the energy at which electrons are injected into the acceleration region $\gamma_0 \sim 0.25 \gamma_R$. Instead of energy dependent escape time-scale, we considered the possibility that the acceleration time scale is energy dependent i.e., $\tau_{acc}= \tau_{acc,R}\left(\frac{\gamma}{\gamma_R}\right)^{\kappa}$ and find that such an Energy Dependent Acceleration (EDA) model gives qualitatively similar results to the EDD one. The spectra are well described by the EDA model and the parameter correlations are roughly as predicted, with  $\xi_{R}^2 \sim 2.89$ keV and  $\gamma_0 \sim 0.19 \gamma_R$. Although the EDD and EDA models successfully represent the spectra and the correlation between fitted parameters are consistent with the models, there are inconsistencies which are discussed below.

For the EDD and EDA models, the best fit values of $\kappa$ ranges from 0.3 -- 1.0. This is somewhat surprising since one would assume that the value of $\kappa$ is determined by micro-physics of the system and at different physical regimes should have discrete values. For example, if the turbulence is of Kolomogrov type its value should be $1/3$,  if it is of Kraichnan type it should be $1/2$ and for Bohm diffusion it should be $1$ \citep{2014ApJ...780...64A, 2019Teraki, 2020Goswami}. One of the above situation should have been applicable to the system giving rise to a constant $\kappa$ or even if the system's micro-physics changed, $\kappa$ should have changed discreetly and perhaps not in a continuous manner as observed. In other words, the models suggest that the turbulence scale index varies rapidly and continuously for these systems. Another issue is the value of $\gamma_R$ as inferred from the correlation between spectral parameters. If the energy dependence of the acceleration or escape time-scale is related to gyration radius of the electron, then $\gamma_R$ may be identified with the energy where the gyration radius becomes as large as the size of the system and hence beyond this energy, the time-scales should be the free streaming values and energy independent.
For the model to be consistent, the typical synchrotron photon energy emitted by particles with  $\gamma_R$, should be much larger than the photon energy range being considered. However, the corresponding photon energy ($\xi^2$) turns out to be within the X-ray spectral range being fitted.  

Perhaps a more generic issue with all such class of models is the variation of the normalization with spectral shape. In general, for these models, there is a continuous injection of electron at a low energy $\gamma_0$ and then the processes of acceleration, diffusion and cooling, create a non-thermal distribution of the electron energies in the steady state. Variation of the injection rate, would lead to a simple up and down scaling of the distribution without changes in its shape. On the other hand, variations in the acceleration or diffusion time-scales would lead to changes in the distribution shape, which in turn would lead to rather dramatic changes in the density of particles at energies $>> \gamma_0$. In the simple picture, where the distribution is a power-law, changes in the index would lead to large variations in the value of the distribution at large $\gamma$ since the power-law would be approximately ``pegged'' at the pivot point $\gamma_0$ \citep{2021MNRAS.504.5485S}. This is generically true and one would expect large variations in the inferred normalization at some large $\gamma$ with spectral shape variations, which is not observed. For the EDD and EDA models considered here, while the observed correlation between the normalization and $\kappa$ can be explained in the framework of the model, the value of $\gamma_0$ turns out to be close to those producing the X-rays ($\gamma_0 \sim 0.2 \gamma_R$), which is physically inconsistent.

The models considered in this work are simple ones which have analytical solutions, but in reality the physical situation may be more complex and a more sophisticated technique to efficiently test non-analytical models need to be developed and applied. For example, both the acceleration and diffusion time-scales could be energy dependent, which would perhaps be more physical \citep{2004Massaro,2006A&A...448..861M,2007Tramacere}. The energy dependence of these time-scale could be different than the power-law form used in this work. Stochastic acceleration (which may be represented as an additional second order differential term in  \autoref{kinetic}) may play an important part in describing the particle distribution.  The observed photon spectrum may be arising from a distribution of particles which having escaped from an acceleration region, and radiate in a surrounding cooling region. Perhaps, an important assumption of these models, that the particle distribution is in steady state may not be valid. In other words, it is implicitly assumed in these models, that the acceleration and escape  time-scales are  significantly smaller than the observed variability time-scale of the photon spectrum. Note that this is not exactly the same as having a one-time injection of particles into the system and studying their evolution but to solve the corresponding time-dependent equation with some physical parameter varying continuously with time. In a recent work \citet{2021MNRAS.504.5485S} have shown that such a model may indeed solve the generic problem of large spectral shape variation with modest changes in the normalization.

In this work, we have shown that time-resolved spectroscopy of {\em{AstroSat}} data of Mkn\,421 can discriminate and rule out simple models. Even though the models considered can explain the spectra of each time segment and even if the correlation between spectral parameters can be explained within the framework of the model, the inferred estimates of physical values can be used to argue against these models in their simplest forms. A primary ingredient to do this is to also consider the variation of the normalization with spectral shape parameters. This opens the need, and provides the motivation to develop more sophisticated models and test them against data.

\section*{Acknowledgements}


We thank the anonymous referee for insightful comments and constructive suggestions. The authors JH and ACP would like to acknowledge Inter University Centre for Astronomy and Astrophysics (IUCAA), Pune, India for providing facilities to carry out this work. ZS is supported by the Department of Science and Technology, Govt. of India, under the INSPIRE Faculty grant (DST/INSPIRE/04/2020/002319).
This publication uses data from the {\em{Astrosat}} mission of the Indian Space Research Organisation(ISRO), archived at the Indian Space Science Data Centre(ISSDC). This work has used the data from the Soft X-ray Telescope (SXT) developed at TIFR, Mumbai. LaxpcSoft software is used for analysis of the LAXPC data and we acknowledge the LAXPC Payload Operation Center (TIFR, Mumbai). This research has made use of data, software and/or web tools obtained from the High Energy Astrophysics Science Archive Research Center (HEASARC), a service of the Astrophysics Science Division
at NASA/GSFC and of the Smithsonian Astrophysical Observatory’s High Energy Astrophysics Division.

\section*{DATA AVAILABILITY}
The data and the model used in this article will be shared on reasonable request to the corresponding author, Jyotishree Hota (email: hotajyoti@gmail.com) or
Zahir Shah (email: shahzahir4@gmail.com).



\bibliographystyle{mnras}
\bibliography{example} 

\begin{table*}
\centering
\footnotesize
\caption{Best fit parameters of time resolved spectral analysis fitted with the synchrotron convolved Log-parabola model and $\xi_{max}$ model.
}
\label{Tab:loggammamax}
\vspace{0.3cm} 
\begin{tabular}{lccccccccccr}
\hline \hline
 Obs. &\multicolumn{5}{c}{Log-parabola} & & \multicolumn{5}{c}{$\xi_{max}$} \\
\cline{2-6} 
\cline{8-12} 

 & $\alpha$ & $\beta$ & norm & $log_{10} F$  & $\chi_{\rm red}^{2}$ & & p & $\xi_{max}$ & norm & $log_{10} F$ & $\chi_{\rm red}^{2}$ \\
\hline
  1 & $2.84_{-0.04}^{+ 0.04}$ & $1.36_{-0.08}^{+0.08}$ & $1.29_{-0.01}^ {+0.01}$ & $-9.076_{-0.004}^{+0.004}$ & 1.13 && $2.69_{-0.15}^{+0.18}$ & $4.80_{-0.15}^{+0.18}$ & $1.39_{-0.01}^{+ 0.01}$ & $-9.063_{- 0.004}^{+0.004}$ & 1.15\\
 2 & $2.98_{-0.03}^{+0.03}$ & $1.11_{-0.06}^{+0.06}$ & $1.33_{-0.01}^{+0.01}$ & $-9.060_{-0.003}^{+0.003}$ & 0.92 && $2.81_{-0.03}^{+0.03}$ & 
  $5.54_{-0.19}^{+0.20}$ & $1.46_{-0.01}^{+0.01}$ & $-9.049_{-0.003}^{+0.003}$ & 0.93\\
  3 & $2.98_{-0.04}^{+0.04}$ & $1.17_{-0.08}^{+0.08}$ & $1.26_{-0.01}^{+0.01}$ & $-9.089_{-0.004}^{+0.004}$ & 0.97 && $2.81_{-0.03}^{+0.03}$ & 
  $5.33_{-0.21}^{+0.23}$ & $1.39_{-0.01}^{+0.01}$ & $-9.079_{-0.004}^{+0.004}$ & 1.02\\
  4 & $3.00_{-0.08}^{+0.08}$ & $0.95_{-0.15}^{+0.15}$ & $1.23_{-0.02}^{+0.02}$ & $-9.083_{-0.008}^{+0.008}$ & 0.97 && $2.85_{-0.07}^{+0.08}$ & $6.04_{-0.56}^{+0.75}$ & $1.35_{-0.03}^{+0.03}$ & $-9.075_{-0.009}^{+0.009}$ & 0.98\\
  5 & $2.92_{-0.05}^{+0.05}$ & $0.99_{-0.08}^{+0.08}$ & $1.31_{-0.01}^{+0.01}$ & $-9.048_{-0.005}^{+0.005}$ & 0.93 && $2.77_{-0.04}^{+0.04}$ & $5.83_{-0.28}^{+0.32}$ & $1.42_{-0.02}^{+0.02}$ & $-9.036_{-0.005}^{+0.005}$ & 0.92\\
  6 & $2.82_{-0.04}^{+0.04}$ & $1.04_{-0.08}^{+0.08}$ & $1.37_{-0.01}^{+0.01}$ & $-9.014_{-0.004}^{+0.004}$ & 1.01 && $2.70_{-0.04}^{+0.04}$ & $5.66_{-0.25}^{+0.27}$ & $1.47_{-0.02}^{+0.02}$ & $-9.002_{-0.004}^{+0.004}$ & 1.02\\
  7 & $2.86_{-0.04}^{+0.04}$ & $1.01_{-0.07}^{+0.07}$ & $1.49_{-0.01}^{+0.01}$ & $-8.982_{-0.004}^{+0.004}$ & 1.00 && $2.73_{-0.03}^{+0.03}$ & $5.77_{-0.24}^{+0.26}$ & $1.60_{-0.02}^{+0.02}$ & $-8.971_{-0.004}^{+0.004}$ & 1.01\\
  8 & $2.81_{-0.04}^{+0.04}$ & $1.26_{-0.08}^{+0.08}$ & $1.54_{-0.01}^{+0.01}$ & $-8.984_{-0.004}^{+0.004}$ & 0.94 && $2.67_{-0.03}^{+0.03}$ & $4.95_{-0.17}^{+0.18}$ & $1.64_{-0.02}^{+0.02}$ & $-8.972_{-0.004}^{+0.004}$ & 0.92\\
  9 & $2.87_{-0.05}^{+0.05}$ & $1.13_{-0.08}^{+0.09}$ & $1.44_{-0.01}^{+0.01}$ & $-9.010_{-0.004}^{+0.004}$ & 0.99 && $2.72_{-0.04}^{+0.04}$ & $5.34_{-0.23}^{+0.25}$ & $1.55_{-0.02}^{+0.02}$ & $-8.999_{-0.005}^{+0.005}$ & 0.99\\
  10 & $2.92_{-0.03}^{+0.03}$ & $1.01_{-0.06}^{+0.06}$ & $1.43_{-0.01}^{+0.01}$ & $-9.009_{-0.003}^{+0.003}$ & 0.93 && $2.77_{-0.03}^{+0.03}$ & $5.72_{-0.22}^{+0.24}$ & $1.56_{-0.01}^{+0.01}$ & $-8.999_{-0.003}^{+0.003}$ & 0.93\\
  11 & $2.89_{-0.04}^{-0.04}$ & $1.03_{-0.07}^{+0.07}$ & $1.48_{-0.01}^{+0.01}$ & $-8.992_{-0.004}^{+0.004} $ & 0.99 && $2.75_{-0.03}^{+0.03}$ & $5.69_{-0.24}^{+0.26}$ & $1.60_{-0.02}^{+0.02}$ & $-8.981_{-0.004}^{+0.004}$ & 0.99\\
  12 & $2.96_{-0.04}^{+0.04}$ & $0.98_{-0.07}^{+0.07}$ & $1.49_{-0.01}^{+0.01}$ & $-8.997_{-0.004}^{+0.004}$ & 0.97 && $2.82_{-0.03}^{+0.03}$ & $5.97_{-0.26}^{+0.29}$ & $1.63_{-0.02}^{+0.02}$ & $-8.987_{-0.004}^{+0.004}$ & 1.00\\
  13 & $2.91_{-0.04}^{+0.04}$ & $1.16_{-0.07}^{+0.08}$ & $1.57_{-0.01}^{+0.01}$ & $-8.983_{-0.004}^{+0.004}$ & 0.98 && $2.76_{-0.03}^{+0.03}$ & $5.33_{-0.21}^{+0.22}$ & $1.71_{-0.02}^{+0.02}$ & $-8.972_{-0.004}^{+0.004}$ & 0.99\\
  14 & $2.85_{-0.05}^{+0.05}$ & $1.31_{-0.09}^{+0.09}$ & $1.60_{-0.02}^{+0.02}$ & $-8.977_{-0.005}^{+0.005}$ & 1.04 && $2.70_{-0.04}^{+0.04}$ & $4.94_{-0.19}^{+0.21}$ &$1.73_{-0.02}^{+0.02}$ & $-8.964_{-0.005}^{+0.005}$ & 1.02\\
  15 & $2.98_{-0.12}^{+0.11}$ & $0.99_{-0.15}^{+0.16}$ & $1.68_{-0.03}^{+0.03}$ & $-8.950_{-0.011}^{+0.010}$ & 0.85 && $2.91_{-0.09}^{+0.10}$ & $6.44_{-0.59}^{+0.78}$ & $1.86_{-0.04}^{+0.04}$ & $-8.945_{-0.011}^{+0.011}$ & 0.88\\
  16 & $2.85_{-0.04}^{+0.04}$ & $1.09_{-0.08}^{+0.09}$ & $1.83_{-0.02}^{+0.02}$ & $-8.894_{-0.004}^{+0.004}$ & 1.00 && $2.69_{-0.03}^{+0.03}$ & $5.25_{-0.23}^{+0.26}$ & $1.97_{-0.02}^{+0.02}$ & $-8.885_{-0.004}^{+0.004}$ & 1.00\\
  17 & $2.79_{-0.04}^{+0.04}$ & $1.09_{-0.06}^{+0.06}$ & $2.00_{-0.01}^{+0.01}$ & $-8.846_{-0.004}^{+0.004}$ & 1.06 && $2.66_{-0.03}^{+0.04}$ & $5.45_{-0.26}^{+0.30}$ & $2.12_{-0.02}^{+0.02}$ & $-8.833_{-0.004}^{+0.004}$ & 1.06\\
  18 & $2.81_{-0.03}^{+0.03}$ & $1.09_{-0.07}^{+0.08}$ & $2.06_{-0.02}^{+0.02}$ & $-8.838_{-0.004}^{+0.004}$ & 1.02 && $2.66_{-0.03}^{+0.03}$ & $5.22_{-0.20}^{+0.22}$ & $2.20_{-0.02}^{+0.02}$ & $-8.828_{-0.004}^{+0.004}$ & 1.00\\
  19 & $2.79_{-0.03}^{+0.03}$ & $1.14_{-0.07}^{+0.07}$ & $2.04_{-0.01}^{+0.01}$ & $-8.841_{-0.003}^{+0.003}$ & 1.08 && $2.66_{-0.03}^{+0.02}$ & $5.29_{-0.19}^{+0.19}$ & $2.17_{-0.02}^{+0.02}$ & $-8.831_{-0.003}^{+0.003}$ & 1.15\\
  20 & $2.67_{-0.03}^{+0.03}$ & $1.38_{-0.06}^{+0.06}$ & $2.22_{-0.01}^{+0.01}$ & $-8.808_{-0.003}^{+0.003}$ & 1.15 && $2.58_{-0.02}^{+0.02}$ & $4.80_{-0.10}^{+0.13}$ & $2.32_{-0.02}^{+0.02}$ & $-8.795_{-0.003}^{+0.003}$ & 1.15\\
  21 & $2.61_{-0.04}^{+0.04}$ & $1.37_{-0.07}^{+0.07}$ & $2.23_{-0.02}^{+0.02}$ & $-8.795_{-0.004}^{+0.004}$ & 0.95 && $2.56_{-0.03}^{+0.03}$ & $4.90_{-0.16}^{+0.13}$ & $2.31_{-0.03}^{+0.03}$ & $-8.782_{-0.004}^{+0.004}$ & 0.99\\
  22 & $2.61_{-0.04}^{+0.04}$ & $1.18_{-0.07}^{+0.07}$ & $2.16_{-0.02}^{+0.02}$ & $-8.787_{-0.004}^{+0.004}$ & 0.94 && $2.57_{-0.03}^{+0.03}$ & $5.37_{-0.18}^{+0.19}$ & $2.25_{-0.03}^{+0.02}$ & $-8.774_{-0.004}^{+0.004}$ & 1.01\\
  23 & $2.54_{-0.03}^{+0.03}$ & $1.09_{-0.06}^{+0.06}$ & $2.22_{-0.02}^{+0.02}$ & $-8.748_{-0.003}^{+0.003}$ & 1.00 && $2.51_{-0.03}^{+0.02}$ & $5.50_{-0.19}^{+0.17}$ & $2.28_{-0.02}^{+0.02}$ & $-8.735_{-0.004}^{+0.003}$ & 1.02\\
  24 & $2.47_{-0.04}^{+0.04}$ & $1.27_{-0.08}^{+0.08}$ & $2.33_{-0.02}^{+0.02}$ & $-8.729_{-0.004}^{+0.004}$ & 1.24 && $2.43_{-0.03}^{+0.02}$ & $4.84_{-0.17}^{+0.13}$ & $2.35_{-0.03}^{+0.03}$ & $-8.716_{-0.004}^{+0.004}$ & 1.23\\
  25 & $2.55_{-0.03}^{+0.03}$ & $1.26_{-0.06}^{+0.07}$ & $2.26_{-0.02}^{+0.02}$ & $-8.760_{-0.003}^{+0.003}$ & 1.20 && $2.49_{-0.02}^{+0.02}$ & $4.86_{-0.15}^{+0.13}$ & $2.31_{-0.02}^{+0.02}$ & $-8.749_{-0.003}^{+0.003}$ & 1.21\\
  26 & $2.47_{-0.03}^{+0.03}$ & $1.59_{-0.07}^{+0.07}$ & $2.25_{-0.02}^{+0.02}$ & $-8.781_{-0.003}^{+0.003}$ & 1.10 && $2.43_{-0.02}^{+0.02}$ & $4.27_{-0.13}^{+0.08}$ & $2.26_{-0.02}^{+0.02}$ & $-8.768_{-0.003}^{+0.003}$ & 1.08\\
  27 & $2.42_{-0.04}^{+0.04}$ & $1.62_{-0.08}^{+0.08}$ & $2.22_{-0.02}^{+0.02}$ & $-8.780_{-0.003}^{+0.003}$ & 1.11 && $2.41_{-0.03}^{+0.02}$ & $4.30_{-0.15}^{+0.08}$ & $2.20_{-0.03}^{+0.03}$ & $-8.765_{-0.004}^{+0.004}$ & 1.13\\
  28 & $2.41_{-0.04}^{+0.04}$ & $1.85_{-0.07}^{+0.07}$ & $2.28_{-0.02}^{+0.02}$ & $-8.790_{-0.003}^{+0.003}$ & 1.07 && $2.42_{-0.02}^{+0.02}$ & $4.13_{-0.03}^{+0.09}$ & $2.27_{-0.02}^{+0.03}$ & $-8.773_{-0.003}^{+0.003}$ & 1.11\\
  29 & $2.35_{-0.05}^{+0.05}$ & $1.71_{-0.08}^{+0.08}$ & $2.29_{-0.02}^{+0.02}$ & $-8.762_{-0.004}^{+0.004}$ & 1.04 && $2.40_{-0.02}^{+0.03}$ & $4.33_{-0.07}^{+0.15}$ & $2.27_{-0.03}^{+0.03}$ & $-8.765_{-0.004}^{+0.004}$ & 1.09\\
  30 & $2.30_{-0.06}^{+0.05}$ & $1.92_{-0.10}^{+0.10}$ & $2.38_{-0.02}^{+0.02}$ & $-8.754_{-0.004}^{+0.004}$ & 1.13 && $2.33_{-0.02}^{+0.04}$ & $3.92_{-0.04}^{+0.18}$ & $2.30_{-0.03}^{+0.04}$ & $-8.745_{-0.004}^{+0.003}$ & 1.12\\
  31 & $2.34_{-0.07}^{+0.07}$ & $2.02_{-0.11}^{+0.11}$ & $2.39_{-0.03}^{+0.03}$ & $-8.770_{-0.005}^{+0.005}$ & 0.99 && $2.37_{-0.04}^{+0.04}$ & $3.91_{-0.11}^{+0.11}$ & $2.34_{-0.05}^{+0.05}$ & $-8.755_{-0.005}^{+0.006}$ & 1.05\\
  32 & $2.26_{-0.04}^{+0.04}$ & $1.94_{-0.09}^{+0.09}$ & $2.28_{-0.02}^{+0.02}$ & $-8.767_{-0.003}^{+0.003}$ & 1.15 && $2.31_{-0.03}^{+0.02}$ & $3.90_{-0.12}^{+0.03}$ & $2.19_{-0.03}^{+0.03}$ & $-8.749_{-0.004}^{+0.002}$ & 1.15\\
  33 & $2.14_{-0.04}^{+0.03}$ & $1.67_{-0.07}^{+0.07}$ & $2.33_{-0.02}^{+0.02}$ & $-8.698_{-0.003}^{+0.003}$ & 1.18 && $2.26_{-0.03}^{+0.01}$ & $4.26_{-0.16}^{+0.04}$ & $2.21_{-0.03}^{+0.02}$ & $-8.684_{-0.003}^{+0.004}$ & 1.19\\
  34 & $2.15_{-0.04}^{+0.04}$ & $1.83_{-0.08}^{+0.08}$ & $2.44_{-0.02}^{+0.02}$ & $-8.700_{-0.003}^{+0.003}$ & 1.19 && $2.26_{-0.03}^{+0.01}$ & $4.09_{-0.18}^{+0.01}$ & $2.31_{-0.03}^{+0.03}$ & $-8.683_{-0.004}^{+0.003}$ & 1.18\\
  35 & $2.16_{-0.04}^{+0.04}$ & $2.33_{-0.07}^{+0.07}$ & $2.42_{-0.02}^{+0.02}$ & $-8.757_{-0.003}^{+0.003}$ & 1.13 && $2.29_{-0.01}^{+0.01}$ & $3.70_{-0.05}^{+0.02}$ & $2.28_{-0.02}^{+0.02}$ & $-8.739_{-0.003}^{+0.002}$ & 1.19\\
  36 & $2.28_{-0.04}^{+0.04}$ & $2.13_{-0.09}^{+0.09}$ & $2.38_{-0.02}^{+0.02}$ & $-8.773_{-0.003}^{+0.003}$ & 0.97 && $2.32_{-0.02}^{+0.02}$ & $3.71_{-0.01}^{+0.07}$ & $2.28_{-0.03}^{+0.03}$ & $-8.757_{-0.003}^{+0.003}$ & 0.96\\
  37 & $2.45_{-0.09}^{+0.08}$ & $1.22_{-0.32}^{+0.36}$ & $2.20_{-0.06}^{+0.07}$ & $-8.745_{-0.013}^{+0.013}$ & 0.97 && $2.33_{-0.02}^{+0.03}$ & $3.54_{-0.03}^{+0.10}$ & $2.34_{-0.05}^{+0.04}$ & $-8.770_{-0.005}^{+0.005}$ & 1.09\\
  38 & $2.51_{-0.09}^{+0.09}$ & $2.05_{-0.19}^{+0.20}$ & $2.29_{-0.04}^{+0.04}$ & $-8.826_{-0.007}^{+0.007}$ & 0.96 && $2.44_{-0.05}^{+0.05}$ & $3.73_{-0.17}^{+0.19}$ & $2.31_{-0.06}^{+0.06}$ & $-8.813_{-0.007}^{+0.007}$ & 0.98\\
  39 & $2.48_{-0.04}^{+0.04}$ & $1.82_{-0.07}^{+0.08}$ & $2.28_{-0.02}^{+0.02}$ & $-8.801_{-0.003}^{+0.003}$ & 1.03 && $2.46_{-0.02}^{+0.02}$ & $4.13_{-0.06}^{+0.10}$ & $2.29_{-0.03}^{+0.03}$ & $-8.787_{-0.003}^{+0.004}$ & 1.08\\
  
\hline \hline
\end{tabular} 
\textbf{Notes}:  $log_{10} F_{0.5-18\,keV}$ is logarithm of flux (ergs cm$^{-2}$ s$^{-1}$) in the range 0.5--18.0 keV.
\end{table*}

\begin{table*}
\centering
\footnotesize
\caption{Best fit parameters of time resolved spectral analysis fitted with the synchrotron convolved energy dependent diffusion model and energy dependent acceleration  model.}
\label{Tab:edaedd}
\vspace{0.3cm} 
\begin{tabular}{lccccccccccr}
\hline \hline
 Obs. &\multicolumn{5}{c}{Energy dependent diffusion (EDD) model} & & \multicolumn{5}{c}{Energy dependent acceleration (EDA) model} \\
\cline{2-6} 
\cline{8-12} 

& $\psi$& $\kappa$ & norm & lg10F & $\chi_{\rm red}^{2}$  & & $\kappa$& $\psi$ & norm & lg10F & $\chi_{\rm red}^{2}$ \\
\hline

 1 & $1.82_{-0.05}^{+0.03}$ & $0.50_{-0.02}^{+0.04}$ & $46.94_{-14.26}^{+13.13}$ & $-9.071_{-0.004}^{+0.004}$ & 1.13 & & $0.44_{-0.03}^{+0.02}$ & $2.24_{-0.02}^{+0.03}$ & $214.30_{-43.70}^{+110.16}$ & $-9.072_{-0.004}^{+0.004}$ & 1.13 \\
  2 & $1.96_{-0.04}^{+ 0.02}$ & $0.40_{-0.02} ^{+ 0.03}$ & $176.34_{-60.39} ^{+ 61.58}$ & $-9.056 _{- 0.003 }^{+ 0.003}$ & 0.92 && $0.35_{-0.02} ^{+ 0.02}$ & $2.31_{- 0.02 }^{+ 0.02}$ & $891.18_{- 245.51}^{+ 447.00}$ & $-9.057 _{- 0.003}^{+ 0.003}$ & 0.92 \\
  3 & $1.96_{-0.05}^{+ 0.03}$ & $0.42_{-0.02}^{+ 0.04}$ & $136.88 _{-51.07}^{+ 53.04}$ & $-9.086_{-0.004}^{+ 0.004}$ & 0.98 && $0.37_{-0.02}^{+ 0.02}$ & $2.33_{-0.02}^{+ 0.02}$ & $707.28_{-235.06}^{+ 347.33}$ & $-9.086 _{- 0.004} ^{+ 0.004}$ & 0.98 \\
  
4 & $1.96_{-0.06}^{+ 0.10}$ & $0.37_{- 0.08}^{+ 0.04}$ & $242.5_{-122.27}^{+ 122.27}$ & $-9.080_{-0.008}^{+0.008}$ & 0.97 && $0.33_{-0.06}^{+  0.03}$ & $2.28_{-0.04}^{+ 0.05}$ & $1335.13_{-676.50}^{+ 676.50}$ & $-9.080 _{-0.008}^{+ 0.008}$ & 0.97 \\
  5 & $1.90_{- 0.05}^{+ 0.02}$ & $0.38_{-0.03} ^{+ 0.04}$ & $201.07_{- 86.05}^{+ 154.30}$ & $-9.043 _{-0.005} ^{+ 0.005}$ & 0.92 && $0.33 _{-0.03}^{+ 0.02}$ & $2.23 _{-0.02}^{+ 0.02}$ & $1042.25 _{- 404.43}^{+ 1021.79}$ & $-9.044 _{-0.005}^{+ 0.005}$ & 0.92 \\
  6 & $1.81 _{-0.05}^{+ 0.03} $& $0.40_{- 0.02}^{+ 0.04}$ & $127.04 _{- 53.44} ^{+56.58}$ & $-9.010_{-    0.004}^{+ 0.004}$ & 1.01 && $0.35 _{-0.02}^{+ 0.03}$ &$2.16_{-0.02}^{+ 0.02}$ & $640.11 _{-237.26}^{+ 372.11}$ & $-9.010_{- 0.004 }^{+ 0.004}$ & 1.01\\
  7 & $1.84 _{- 0.04 }^{+ 0.03}$ & $0.39 _{-0.03}^{+ 0.03}$ & $163.54 _{-58.98} ^{+ 84.97} $& $-8.978_{-0.004 }^{+ 0.004}$ & 0.99 && $0.34 _{- 0.02}^{+ 0.02} $ & $2.18 _{- 0.02 }^{+ 0.02} $ & $ 875.75 _{-302.27}^{+ 519.84}$ & $-8.979 _{- 0.004 }^{+ 0.004 }$ & 0.99\\
  8 & $1.79 _{- 0.05}^{+ 0.02}$ & $0.48 _{- 0.02}^{+ 0.04}$ & $63.28 _{- 19.40}^{+ 18.48}$ & $-8.980 _{-0.004}^{+ 0.004}$ & 0.93 && $0.42 _{- 0.03}^{+ 0.02}$ & $2.19 _{- 0.02}^{+ 0.02} $& $294.60 _{-61.89}^{+ 153.88}$ & $-8.980 _{- 0.004}^{+ 0.004}$ & 0.93\\
  9 & $1.85 _{-0.05}^{+ 0.04}$ & $0.43 _{-0.03}^{+ 0.04}$ & $103.14 _{-36.55}^{+ 55.09}$ & $-9.006 _{-0.004}^{+ 0.004}$ & 0.99 && $0.38 _{- 0.03}^{+ 0.02}$ & $2.21_{- 0.02}^{+ 0.03}$ & $498.99 _{- 142.52}^{+ 398.01}$ & $-9.006_{- 0.004}^{+ 0.004}$ & 0.99\\
  10 & $1.90 _{-0.03}^{+ 0.03} $ & $ 0.38 _{- 0.02}^{+ 0.03} $ & $204.21_{- 66.56} ^{+ 94.64}$ & $-9.006 _{- 0.003} ^{+ 0.003}$ & 0.93 && $0.34 _{- 0.02 }^{+ 0.02}$ & $2.24 _{- 0.02}^{+ 0.02} $ & $1085.63 _{-318.21}^{+ 625.98}$ & $-9.006 _{- 0.003}^{+ 0.003}$ & 0.93\\
  11 & $1.88 _{-0.04}^{+ 0.04} $ & $0.39 _{- 0.03}^{+ 0.03}$ & $177.95 _{- 65.98}^{+ 98.79}$ & $-8.988 _{-0.004}^{+ 0.004}$  & 0.98 && $0.34 _{- 0.02}^{+ 0.02} $& $2.22 _{- 0.02}^{+ 0.02}$ & $939.96_{-321.74} ^{+ 631.42}$ & $-8.988 _{- 0.004 }^{+ 0.004} $& 0.98\\
  12 & $1.95 _{- 0.04} ^ {+0.03} $& $0.36 _{- 0.03}^{+ 0.03} $ & $319.35 _{- 122.48}^{+ 204.53}$ & $-8.993 _{- 0.004}^{+ 0.004} $& 0.97 & & $0.32 _{-0.02} ^{+ 0.02}$ &$ 2.27_{- 0.02}^{+ 0.02 }$& $1789.81 _{-689.15 }^{+ 1224.59} $&$ -8.994 _{- 0.004}^{+ 0.004} $& 0.97\\
  13 & $1.89 _{- 0.04}^{+ 0.04} $& $0.43_{- 0.03}^{+ 0.03} $& $124.96 _{-39.96}^{+ 59.58} $& $-8.979 _{- 0.004}^{+ 0.004} $& 0.98 & & $0.38 _{- 0.03}^{+ 0.02} $&$ 2.26 _{- 0.02}^{+ 0.02} $& $619.33 _{- 161.62}^ {+424.85} $& $-8.980 _{- 0.004} ^{+ 0.004} $& 0.98\\
  14 & $1.82 _ {-0.05} ^{+ 0.05}$ & $0.49 _{ -0.04 }^{+ 0.04}$ & $61.95 _{-18.57 }^{+ 29.53}$ & $-8.972 _{- 0.005} ^{+ 0.005}$ & 1.03 && $0.42 _{- 0.03}^{+ 0.02 }$& $2.23 _{- 0.03}^{+ 0.03} $& $319.51 _{- 91.67} ^{+ 184.50 }$& $-8.973 _ {-0.005 }^{+0.005}$ & 1.03\\
  15 & $1.95 _{-0.08} ^{+ 0.11}$ & $0.38_{- 0.08}^{+ 0.05}$ & $290.08 _{- 159.55}^{+ 159.55}$ & $-8.947 _{- 0.011} ^ {+0.011} $& 0.85 && $0.33 _{- 0.07}^{+ 0.04}$ & $2.27 _{- 0.04}^{+ 0.06}$ & $1519.87 _{- 830.64 }^{+ 830.64} $& $-8.947 _{- 0.011}^{+ 0.011}$ & 0.85\\
  16 & $1.82_{-0.06}^{+0.03}$ & $0.43_{-0.02}^{+0.05}$ & $128.74_{-53.95}^{+48.71}$ & $-8.892_{-0.004}^{+0.004} $& 1.00 && $0.38_{-0.03}^{+0.02} $& $2.18_{-0.02}^{+0.02}$ & $577.66_{-166.21}^{+399.66} $& $-8.892_{-0.004}^{+0.004} $ & 1.00\\
  17 & $1.77_{-0.05}^{+0.03}$ & $0.41_{-0.03}^{+0.05} $& $140.72_{-59.26}^{+62.65}$ & $-8.840_{-0.004}^{+0.004}$ & 1.05 & & $0.37_{-0.03}^{+0.02}$ & $2.12_{-0.02}^{+0.02} $& $614.09_{-175.44}^{+514.86} $& $-8.841_{-0.004}^{+0.004} $& 1.07\\
  18 & $1.78_{-0.05}^{+0.03}$ & $0.44_{-0.02}^{+0.04} $& $119.11_{-44.37}^{+35.35}$ & $-8.835_{-0.004}^{+0.004}$ & 1.00 && $ 0.38_{-0.02}^{+0.02} $&$ 2.15_{-0.02}^{+0.02}$ & $579.50_{-168.66}^{-259.54}$ & $-8.836_{-0.004}^{+0.004}$ & 1.01\\
  19 & $1.76_{-0.03}^{+0.02}$ & $0.45_{-0.02}^{+0.03} $& $100.68_{-28.84}^{+26.77}$ & $-8.838_{-0.003}^{+0.003}$ & 1.09 && $0.39_{-0.02}^{+0.02} $ & $ 2.14_{-0.02}^{+0.02}$ & $493.87_{-113.25}^{+189.08}$ & $-8.839_{-0.003}^{+0.003}$ & 1.09\\
  20 & $1.65_{-0.03}^{+0.03}$ & $0.55_{-0.03}^{+0.03}$ & $43.84_{-7.74}^{+9.94}$ & $-8.804_{-0.003}^{+0.003}$ & 1.14 & & $0.45_{-0.02}^{+0.02} $&   $ 2.10_{-0.02}^{+0.02}$ & $224.03_{-41.00}^{+55.94}$ & $-8.804_{-0.003}^{+0.003}$ & 1.14\\
  21 & $1.58_{-0.03}^{+0.06}$ & $0.56_{-0.04}^{+0.02}$ & $36.29_{-5.09}^{+13.56}$ & $-8.790_{-0.004}^{+0.004}$ & 0.95 && $0.46_{-0.02}^{+0.02}$ & $2.05_{-0.01}^{+0.02}$ & $195.89_{-39.26}^{+55.69}$ & $-8.791_{-0.004}^{+0.004}$ & 0.95\\
  22 & $1.60_{-0.04}^{+0.04}$ & $0.48_{-0.03}^{+0.03}$ & $58.37_{-14.23}^{+20.59}$ & $-8.782_{-0.004}^{+0.004}$ & 0.95 && $ 0.41_{-0.02}^{+0.02}$ & $2.01_{-0.02}^{+0.02}$ & $287.94_{-67.12}^{+125.10}$ & $-8.783_{-0.004}^{+0.004}$ & 0.95\\
  23 & $1.53_{-0.03}^{+0.03} $& $0.47_{-0.03}^{+0.03}$ & $ 55.11_{-11.75}^{+16.30} $ & $-8.743_{-0.003}^{+0.003} $ & 0.99 && $ 0.40_{-0.02}^{+0.02}$ & $1.92_{-0.02}^{+0.02} $& $273.81_{-55.50}^{+101.95}$ & $-8.744_{-0.003}^{+0.003}$ & 0.99\\
  24 & $1.43_{-0.04}^{+0.04} $& $0.58_{-0.04}^{+0.04} $& $27.10_{-5.30}^{+7.31} $& $ -8.725_{-0.004}^{+0.004} $ & 1.22 & & $ 0.47_{-0.03}^{+0.02} $ & $1.89_{-0.02}^{+0.02} $ & $ 127.10_{-21.97}^{+47.54} $ & $-8.726_{-0.004}^{+0.004} $ & 1.23\\
  25 & $1.51_{-0.03}^{+0.03}$ & $0.55_{-0.03}^{+0.03} $& $34.16_{-5.79}^{+7.65}$ & $-8.757_{-0.003}^{+0.003}$ & 1.19 & & $0.45_{-0.02}^{+0.02} $ & $ 1.96_{-0.01}^{+0.02} $& $166.69_{-25.87}^{+48.02}$ & $-8.758_{-0.003}^{+0.003} $ & 1.19\\
  26 & $1.43_{-0.03}^{+0.03} $ & $ 0.69_{-0.03}^{+0.04} $ & $ 17.18_{-2.31}^{+2.58} $ & $-8.777_{-0.003}^{+0.003} $ & 1.08 && $ 0.55_{-0.02}^{+0.02}$ &$ 1.97_{-0.02}^{+0.02}$ & $80.94_{-10.58}^{+15.46} $ & $ -8.778_{-0.003}^{+0.003}$ & 1.08\\
  27 & $1.39_{-0.05}^{+0.03} $& $ 0.70_{-0.03}^{+0.04}$ & $15.46_{-2.51}^{+2.20}$  & $ -8.775_{-0.003}^{+0.003} $ & 1.10 & & $0.56_{-0.03}^{+0.02} $ & $1.93_{-0.02}^{+0.02}$ & $68.19_{-8.36}^{+16.21}$ & $ -8.776_{-0.003}^{+0.003} $& 1.10\\
  28 & $1.39_{-0.04}^{+0.04} $& $0.76_{-0.03}^{+0.04} $& $ 13.70_{-1.70}^{+1.89} $ & $ -8.783_{-0.003}^{+0.003}$  &  1.05 && $0.59_{-0.02}^{+0.02} $& $1.98_{-0.02}^{+0.02}$ & $62.28_{-7.70}^{+10.53}$ & $ -8.784_{-0.003}^{+0.003}$ & 1.05\\
  29 & $ 1.35_{-0.05}^{+0.04} $ & $ 0.73_{-0.04}^{+0.04} $ & $ 14.19_{-2.25}^{+2.54} $ & $ -8.756_{-0.004}^{+0.004} $ & 1.03 && $ 0.56_{-0.02}^{+0.03} $ & $ 1.92_{-0.03}^{+0.02} $ &  $67.99_{-14.43}^{+10.59} $ & $ -8.757_{-0.004}^{+0.004} $ & 1.03\\
  30 & $ 1.27_{-0.06}^{+0.05} $ &  $ 0.84_{-0.05}^{+0.05} $ & $ 10.16_{-1.45}^{+1.67} $ &  $ -8.748_{-0.004}^{+0.004}$ & 1.11 && $ 0.64_{-0.03}^{+0.03} $ & $ 1.90_{-0.03}^{+0.03} $ & $44.68_{-6.62}^{+9.24}$ & $ -8.750_{-0.004}^{+0.004} $ & 1.11\\
  31 & $ 1.32_{-0.07}^{+0.05} $ & $ 0.84_{-0.05}^{+0.07} $ & $ 10.92_{-2.02}^{+1.91} $ & $ -8.764_{-0.005}^{+0.005}$ & 0.99 && $ 0.65_{-0.04}^{+0.03}$ & $1.95_{-0.03}^{+0.04} $&  $45.17_{-6.74}^{+13.14}$ & $-8.765_{-0.005}^{+0.005}$ & 1.00\\
  32 & $ 1.23_{-0.04}^{+0.04} $ & $0.87_{-0.04}^{+0.05} $ &  $8.86_{-0.97}^{+1.09} $&$ -8.762_{-0.003}^{+0.003} $ & 1.13 && $0.65_{-0.03}^{+0.03}$ & $ 1.88_{-0.02}^{+0.02} $& $38.74_{-4.49}^{+5.89}$ & $-8.763_{-0.003}^{+0.003}$ & 1.13\\
  33 & $1.12_{-0.04}^{+0.03}$ &$ 0.82_{-0.03}^{+0.04} $ & $8.82_{-0.90}^{+0.95}$ &$ -8.693_{-0.003}^{+0.003} $& 1.16& & $0.61_{-0.02}^{+0.02}$ & $1.72_{-0.02}^{+0.02} $& $ 37.91_{-3.96}^{+5.23} $ & $ -8.694_{-0.003}^{+0.003}$ & 1.18\\
  34 & $ 1.13_{-0.04}^{+0.04} $ & $0.87_{-0.04}^{+0.05} $ & $ 8.51_{-0.92}^{+0.98}$ &  $-8.695_{-0.003}^{+0.003} $ & 1.16 && $0.64_{-0.02}^{+0.03} $ & $ 1.77_{-0.02}^{+0.02} $ & $ 36.69_{-4.51}^{+4.97}$ & $ -8.696_{-0.003}^{+0.003} $ & 1.17\\
  35 & $1.16_{-0.03}^{+0.03}$ &$ 1.00_{-0.04}^{+0.04} $& $7.24_{-0.53}^{+0.60} $& $-8.751_{-0.003}^{+0.003} $ & 1.11 && $ 0.73_{-0.02}^{+0.02} $ & $1.88_{-0.02}^{+0.02} $ & $ 30.37_{-2.60}^{+2.94} $ & $ -8.752_{-0.003}^{+0.003} $ & 1.11\\
  36 & $1.26_{-0.05}^{+0.04} $& $0.91_{-0.04}^{+0.04}$ & $9.01_{-0.92}^{+0.98}$ & $-8.767_{-0.003}^{+0.003}$ & 0.94 & & $0.68_{-0.02}^{+0.03} $& $1.94_{-0.02}^{+0.02} $& $39.81_{-4.59}^{+4.97} $ & $-8.768_{-0.003}^{+0.003} $& 0.94\\
  37 & $1.24_{-0.06}^{+0.06} $& $1.02_{-0.06}^{+0.06}$ & $7.74_{-0.94}^{+1.14} $& $-8.780_{-0.005}^{+0.005} $& 0.97 && $0.75_{-0.03}^{+0.03}$ & $1.98_{-0.03}^{+0.03} $& $32.41_{-4.51}^{+5.62}$ & $-8.781_{-0.005}^{+0.005}$ & 0.96\\
  38 & $1.49_{-0.10}^{+0.08} $& $0.81_{-0.07}^{+0.10}$ & $13.88_{-3.73}^{+4.36}$ & $-8.821_{-0.007}^{+0.007}$ & 0.96 && $0.62_{-0.04}^{+0.07}$ & $2.12_{-0.05}^{+0.04} $& $67.46_{-21.68}^{+21.47}$ & $-8.823_{-0.007}^{+0.007} $& 0.96\\
  39 & $1.47_{-0.05}^{+0.03}$ & $0.73_{-0.03}^{+0.04}$ & $16.42_{-2.64}^{+2.18}$ & $-8.797_{-0.003}^{+0.003}$ & 1.03 && $0.58_{-0.03}^{+0.02}$ & $2.03_{-0.02}^{+0.03}$ & $71.91_{-8.22}^{+17.04}$ & $-8.797_{-0.003}^{+0.003}$ & 1.03\\

\hline \hline
\end{tabular} 
\textbf{Notes}:  $log_{10} F_{0.5-18\,keV}$ is logarithm of flux (ergs cm$^{-2}$ s$^{-1}$) in the range 0.5--18.0 keV.
\end{table*}






\bsp	
\label{lastpage}
\end{document}